\documentclass{aa}  
\usepackage{array}
\newcolumntype{C}[1]{>{\centering\arraybackslash}p{#1}}
\usepackage{graphicx, bm}
\graphicspath{{./fig/}{./png/}}
\usepackage{txfonts}
\usepackage{lipsum}
\usepackage[colorlinks=true,linkcolor=blue,citecolor=blue]{hyperref}
\usepackage{subcaption}         
\usepackage{lscape}             
\usepackage{placeins}           
                                
\usepackage{cleveref}
\usepackage[dvipsnames]{xcolor} 
\bibpunct{(}{)}{;}{a}{}{,}

\usepackage{xcolor}

\newcommand{\fff}{{\bm f}}

\newcommand{\AAA}{{\bm A}}
\newcommand{\mAAA}{\overline{\bm A}}

\newcommand{\BBB}{{\bm B}}
\newcommand{\mBBB}{\overline{\bm B}}

\newcommand{\JJJ}{{\bm J}}
\newcommand{\mJJJ}{\overline{\bm J}}

\newcommand{\mUUU}{\overline{\bm U}}

\newcommand{\gggg}{{\bm g}}
\newcommand{\mEMF}{\overline{\bm{\mathcal{E}}}}

\newcommand{\SSS}{\bm{\mathsf{S}}}
\newcommand{\EQ}{\begin{equation}}
\newcommand{\EN}{\end{equation}}
\newcommand{\EQA}{\begin{eqnarray}}
\newcommand{\ENA}{\end{eqnarray}}

\newcommand{\etaT}{\eta_{\rm T}}

\newcommand{\Prot}{P_{\rm rot}}
\newcommand{\Beq}{B_{\rm eq}}

\newcommand{\Co}{{\rm Co}}

\newcommand{\Costar}{{\rm Co}_\star}

\newcommand{\PraSGS}{{\rm Pr}_{\rm SGS}}

\newcommand{\PrM}{{\rm Pr}_{\rm M}}

\newcommand{\Rey}{{\rm Re}}

\newcommand{\Ros}{{\rm Ro}}

\newcommand{\ReM}{{\rm Re}_{\rm M}}
\newcommand{\Ro}{{\rm Ro}}

\newcommand{\Ta}{{\rm Ta}}

{}

\newcommand{\Lratio}{L_{\rm ratio}}
\def\onethird{{\textstyle{1\over3}}}

\definecolor{ForestGreen}{RGB}{34,139,34}
\definecolor{AGray}{rgb}{.4,.4,.4}
\definecolor{LightYellow}{rgb}{1.,1.,.8}
\definecolor{LightCyan}{rgb}{0.88,1,1}

\begin{document}

   \title{Mean-field interpretation of star-in-a-box simulations of red giants}



   \author{C. A. Ortiz-Rodríguez\inst{1} \corrauth{carolina.ortiz.rodriguez@uni-hamburg.de}
        \and A. Brandenburg \inst{2,3} \email{brandenb@nordita.org}
        \and P. J. K\"apyl\"a \inst{4,5} \email{pkapyla@leibniz-kis.de}
        }

   \institute{Hamburger Sternwarte, Universit\"at Hamburg, Gojenbergsweg 112, 21029 Hamburg, Germany\\
            \email{carolina.ortiz.rodriguez@uni-hamburg.de}
            \and Nordita, KTH Royal Institute of Technology and Stockholm University, Hannes Alfv\'ens v\"ag 12, 10691 Stockholm, Sweden 
            \and The Oskar Klein Centre, Department of Astronomy, Stockholm University, AlbaNova, 10691 Stockholm, Sweden
            \and Institut f\"ur Sonnenphysik (KIS), Georges-K\"ohler-Alle 401a, 79110 Freiburg im Breisgau, Germany
            \and
            Institute of Physics, University of Graz, 8010 Graz, Austria}

   \date{}

 
  \abstract
  {The origin of magnetic fields and the dynamo mechanism in red giants are still not fully understood.}
  {We aim to interpret the dynamo behaviour of global 3D simulations of red giants using mean-field dynamo models.} 
   {We use mean-field models constrained by the differential rotation profile extracted from 3D simulations.
    We perform $\alpha^2$, $\alpha\Omega$, and $\alpha^2\Omega$ mean-field dynamo simulations, varying the strength of the $\alpha$ effect and the differential rotation.}
   {The mean-field models can reproduce the growth rate of several 3D runs.
   The morphology of the large-scale magnetic field is better
   reproduced for the slowly rotating 3D cases than for the rapidly
   rotating ones.
   Faster rotation enhances dynamo action and it also modifies the
   dominant mode of the dynamo. Rapidly rotating 3D runs produce
   predominantly non-axisymmetric equatorial dipoles instead of
   axisymmetric fields at slower rotation.
   Mean-field models are supercritical even in the absence of
   differential rotation, indicating $\alpha^2$ dynamo action.
   By contrast, models using the $\alpha\Omega$ approximation 
   require sufficiently strong differential rotation to become 
   supercritical.
   } 
   {Our results suggest that the magnetic field in red giants requires dynamo action.
   In our mean-field runs, differential rotation speeds up the magnetic decay,
   disfavouring the idea of a persistent fossil magnetic field in red giants
   {as the explanation for the observed surface fields.}}


   \maketitle

\nolinenumbers
\section{Introduction}
Magnetic fields {at the surface of} red giants have been measured using different methods
such as chromospheric emission lines {of \textsc{Ca}\,\textsc{ii}\ H \& K} \citep[e.g.][]{schroder2018stellar} and 
Zeeman splitting \citep[e.g.][]{auriere2015magnetic}. The origin of the
magnetism is likely a dynamo in the deep convective envelopes of these
stars. The interior structure of red giants resembles qualitatively
that of late-type main sequence stars and therefore
magnetic fields may be generated by an $\alpha \Omega$ or
$\alpha^2\Omega$ dynamo which are often used to explain the solar
dynamo \citep[e.g.][and references
therein]{Ossendrijver_2003_AARv_11_287}. Using
stellar structure models, \cite{charbonnel2017magnetic} suggested that an
$\alpha \Omega$ dynamo operates in the convective envelope of low- and 
intermediate mass stars at particular evolutionary phases: the first 
dredge-up, which occurs at the base of the Red Giant Branch (RGB) during 
core-helium burning, and the early Asymptotic Giant Branch (AGB).

Using Zeeman Doppler imaging, \cite{auriere2011} reported that the 
slowly rotating G8 sub-giant EK Eridani has a dominant dipolar 
large-scale magnetic field. They suggested that the magnetic properties of EK
Eridani are the result of the interactions of the remnant of an Ap star and 
the deep convection zone. \cite{Bonanno+19} studied the origin of the
magnetic field in EK Eri.
They used a mean-field model with radially decreasing angular velocity
and found the critical rotation period below dynamo action 
is possible, and
argued that an $\alpha^{2}\Omega$ dynamo or any dynamo including
sufficiently strong $\alpha$ effect
as the possible origin for the observed magnetic field.
A dipolar surface magnetic field was also detected in the giant $\beta$ Ceti 
by \cite{tsvetkova2013magnetic}. 
Given its long rotation period of about 215 days and a Rossby
number $\Ro$ (the ratio of the rotation period $\Prot$ and the
convective turnover time $\tau_c$) of about 1.3, the observed level
of magnetic activity is unexpectedly high, and no significant
surface differential rotation has been measured. They concluded that
dynamo action alone cannot be responsible for explaining the magnetic activity 
of $\beta$ Ceti.
Instead, they suggest that the star may be the descendant of an 
Ap star, exhibiting a remnant of its magnetic field. Giants with a more 
complex surface magnetic field are also known. V390 Aurigae with a 
rotation period of about $9.8$ days shows a complex surface magnetic 
field, with an indication of a toroidal field 
\citep{konstantinova2012magnetic}.
{These authors estimated the Rossby number to be between
0.08 and 0.19 in the lower half of the convective zone and
suggested this low value to} indicate
that an $\alpha\Omega$ dynamo can explain these magnetic fields.
These {studies show} that both 
dynamo-generated and fossil magnetic fields may play a role in 
{the surface magnetism of}
evolved stars. 
{Nevertheless, it is crucial to distinguish between
dynamo-generated magnetic fields in the convective envelopes 
of red giants and the magnetic fields of fossil origin
in the radiative cores, as recent asteroseismic studies have 
begun to provide constraints on the magnetic fields in 
the cores of red giants \citep{fuller2015asteroseismology,bugnet_l_magnetic_2021,li_magnetic_2022,hatt_asteroseismic_2024}.
}

Beyond individual stars, magnetic activity 
in main sequence and evolved stars has also been investigated in a broader statistical context.
For main-sequence stars, the rotation-activity relation has been studied 
thoroughly for both partially and fully convective stars. 
\cite{wright2018stellar} showed that the fractional X-ray luminosity 
$L_X / L_{\mathrm{bol}}$, a proxy of magnetic activity, follows 
a relation with the Rossby number $\Ros$.
In the unsaturated regime ($\Ros \geq 0.1$), the X-ray luminosity
decreases with increasing $\Ros$, while for lower values it reaches a 
saturation level of $L_X/L_{\rm{bol}} \sim
10^{-3}$. {\cite{noyes1984rotation} found that the mean
  chromospheric flux ratio of \textsc{Ca}\,\textsc{ii}\ H \& K lines
  decreases in the unsaturated regime in anti-correlation with $\Ro$.  }
{Moreover, chromospheric 
emission suggests that evolved slowly rotating late-type stars follow a similar 
trend with $\Ro$, suggesting
that turbulence is
important for magnetic fields in main-sequence and evolved
late-type stars \citep{lehtinen2020common}. 
}
{Another proxy to characterise magnetic activity is the photospheric magnetic
index $S_{\rm ph}$, which has been used for F main sequence stars 
\citep{mathur_magnetic_2014,mathur2025magnetic}, although these data show more scatter than the chromospheric index.
}

An important constraint on the internal dynamics of red giants comes from
asteroseismology, which shows that their cores 
rotate faster than their envelopes. Early studies using Kepler data
and theoretical models concluded that the core of the star KIC 8366239
must rotate at least 
ten times faster than its envelope \citep{beck2012fast}.
Additionally, \cite{deheuvels2012seismic} showed that
an early red giant star core must rotate at least five times faster than
its envelope. More recent analysis of larger samples
\citep{li2024asteroseismic} report typical core-to-envelope rotation
rate ratio of about $20$, although a subset of stars exhibit much weaker 
differential rotation, with the core rotating less than twice as fast as the envelope.
The origin of this core-to-envelope rotation ratio is not fully understood.
As the star evolves, its envelope expands and the core contracts, leading the core
to speed up due to the conservation of angular momentum. However,
according to some studies
\citep{cantiello_angular_2014, aerts2019angular}, the cores of
many observed red giants rotate slower than expected{, suggesting
that angular momentum is extracted from the core.
The interaction between mixed modes is a
potential mechanism to redistribute angular momentum, and has been
found to play an important role in the most evolved low-mass stars
\citep{Belkacem2015}. However, the importance of mixed modes is more
limited in the subgiant and early red giant
phases. \cite{bordadagua_efficiency_2025}
found that mixed modes alone cannot comprehensively explain the rotation rates 
observed in red giants. Magnetic fields,
meridional circulation, and hydrodynamic instabilities should be taken into 
consideration \citep{aerts2019angular}.
Another mechanism
}
that redistributes angular momentum is the $\Lambda$ effect, which is the non-diffusive
part of the Reynolds stress tensor, generated by rotating, anisotropic convection
\citep{Ruediger_1989,kapyla2008lambda,Kapyla_2026}.
Numerical simulations of red giants support this picture. For example, 
\cite{Brun2009} performed 3D anelastic hydrodynamical simulations of
the inner envelope of a rotating red giant, finding shellular and cylindrical differential
rotation profiles depending on the Rossby number, while maintaining large radial shear.
These results highlight the role of rotating convection in shaping the internal
rotation of red giants.
Numerical simulations therefore play an important role in interpreting these 
observations and in understanding the mechanism of the generation and 
maintenance of magnetic fields in red giants. As a first approach,
nonrotating hydrodynamic simulations have been reported in the literature. 
\cite{freytag2002spots} used
3D star-in-a-box simulations with the CO$^5$BOLD code \citep{CO5BOLD2012} of the red supergiant
Betelgeuse, finding giant convective cells dominating its surface. 
Later, \cite{Dorch_2004_AA_423_1101} used the \textsc{Pencil Code} 
\citep{brandenburg2021pencil} to 
perform magnetohydrodynamic simulations with the stellar 
parameters of Betelgeuse, to study whether magnetic fields exist
in such model. This simulation covered the kinematic growth and
non-linear saturation of the dynamo with magnetic fields saturating
slightly above equipartition. This model did not include rotation 
and the magnetic field was due to a small-scale dynamo. The presence of
magnetic fields on the surface
of Betelgeuse was reported observationally for the first time by \cite{auriere2010magnetic}.

More recently, \cite{Amard_et_al_2024_ApJ_974_311} performed 
magnetohydrodynamic simulations of the envelope of a red giant star
with the parameters of Pollux. Their models 
focused on varying the Prandtl number and the seed magnetic field. 
They found a correlation between the size of the convective cells and the 
geometry of the magnetic field, with larger cells leading to stronger 
large-scale magnetic field. For one
of their models, the mean azimuthal field reached $2\,$G near the 
outer radius. This is similar to the
observed surface field strength of Pollux ($0.1$--$0.7\,{\rm G}$).
These 3D simulations of MHD turbulence in red giants provide
a comprehensive description of convection and magnetic fields.
They have shown both similarities and differences compared with the
magnetic field evolution of main sequence stars.
For example, both show large-scale magnetic fields that exhibit a similar
spatio-temporal evolution.
For main sequence stars, and especially for the Sun, however, the
theoretical understanding of the latitudinal evolution is still incomplete.
In particular, the role of meridional circulation is unclear.
For the Sun, it has been proposed that the equatorial return flow at the
bottom of the convection zone is directly responsible for the equatorward
migration of the magnetic field \citep{Choudhuri+95}.
The conventional mean-field dynamo approach, by contrast, explains the
equatorward migration as the result of a dynamo wave that travels to
the equator when the differential rotation has a negative radial gradient 
\citep{parker1955hydromagnetic}.
To shed light on this puzzle, one may hope that the study of the
mean-field evolution in red giants can yield useful clues.
The current paper is a first study in this direction{, focusing
specifically on the dynamo mechanism and differential rotation operating
within the convective envelopes of red giants.}
\section{Models}
\subsection{Star-in-a-box setup}
The 3D model was adapted by Ortiz-Rodriguez et al (in prep.), based on
\cite{dobler2006magnetic} and \cite{kapyla2021star}, in which the star
of radius $R$ is embedded in a {Cartesian domain} of side $2.2R$.
We use the \textsc{Pencil Code} \citep{brandenburg2021pencil}, which solves the
equations of magnetohydrodynamics given by
\begin{align}
    \frac{\partial \AAA}{\partial t} &= \bm{u} \times \BBB - \eta \mu_0 \JJJ,
    \\
    \frac{{\rm D} \ln \rho}{{\rm D} t} &= -\bm{\nabla} \bm\cdot \bm{u},
    \\
    \frac{{\rm D} \bm{u}}{{\rm D} t}  &= -\bm{\nabla} \Phi\!-\!\frac{1}{\rho} 
    \left(\bm{\nabla}p\!-\!\bm{\nabla}\!\bm\cdot\!2 \nu \rho\SSS \!+\!\JJJ
    \times \BBB \right)-2 \bm{\Omega} \times \bm{u}\!+\!\bm{f}_d,
    \\
    \rho T \frac{{\rm D} s}{{\rm D} t} &= \mathcal{H - C} - \bm{\nabla}\bm\cdot
    \left(\bm{F}_{\rm rad}\!+\!\bm{F}_{\rm SGS} \right)\!+\! \mu_0 \eta \JJJ^2
    + 2\rho \nu\SSS^2,
\end{align}
where $\AAA$ is the magnetic vector potential, $\bm{u}$ is the velocity field,
$\BBB = \nabla \times \AAA$ is the magnetic field, $\eta$ is the magnetic 
diffusivity, $\mu_0$ is the magnetic permeability of vacuum, $\JJJ$ is the current
density, $D/Dt = \partial/\partial t + \bm{u}\bm\cdot \nabla$ is the advective 
derivative, $\rho$ is the density of the fluid,
$\Phi$ is the gravitational potential given in terms of the gravitational acceleration
as $\gggg = - \bm\nabla \Phi$ with $\gggg = g \hat{\bm{e}}_r$, given by \cite{freytag2002spots},
\begin{equation}
   g(r) =  -GM \frac{ b^{3n}r^{n-1}\left[ a^n \, \left(b^{2n} + r^{2n}\right)^{0.5} + b^n \, r^n \right]^{(-1-n)/n}}   { \left( b^{2n} + r^{2n} \right)^{(-1+2n)/2n}}.
\end{equation}
Here, $G$ is the gravitational constant, $M$ is the mass of the star,
$a=0.1$, $b=1.0$, and $n=4$. Furthermore, $p$ is the pressure, $\nu$ is the
kinematic viscosity, $\SSS$ is the traceless rate-of-strain tensor,
$\bm\Omega = (0,0,\Omega_0)$ is the rotation rate of the star,
$\fff_d$ describes damping of velocity outside of the star,
$s$ is the specific entropy, $\mathcal{H}$ and
$\mathcal{C}$ are heating and cooling functions, respectively.
Heating is a parameterisation to mimic nuclear energy production in the core
of the star, while cooling removes energy near the surface.
$\bm{F}_{\rm rad} = - K \bm\nabla T$ is the radiative flux, where $K$ is
based on the Kramers opacity law.
$\bm{F}_{\rm SGS} = -\chi_{\rm SGS} \rho \bm\nabla s^\prime$ is the
sub-grid scale flux, which damps fluctuations near grid scale.
Here, $s^\prime$ are the fluctuations of entropy computed
as the difference between the instantaneous and time-averaged specific
entropy; see \cite{kapyla2021star} for a detailed description.
{The star-in-a-box setup avoids coordinate singularities on the
  axis and at the centre of the star. Furthermore, no explicit
  boundary conditions are applied at the surface of the star, which
  allows the magnetic field to evolve more freely.}
\subsubsection{{Stellar parameters}}
{We obtained the stellar parameters using Modules for Experiments in 
Stellar Astrophysics
\citep[MESA,][]{2011ApJS..192....3P, 2013ApJS..208....4P, 2015ApJS..220...15P, 2018ApJS..234...34P, 2019ApJS..243...10P, 2023ApJS..265...15J} code.
Our model corresponds to a red giant at the bottom of the RGB with 
mass, radius, luminosity, effective temperature, and central density 
$M=0.99\,M_{\odot}$, $R=3.1\,R_{\odot}$, $L=4.8\,L_{\odot}$, $T=4830\,K$, 
$\rho_0 = 6.25\times10^7\,\rm{kg/m^3}$,
respectively. The subscript $\odot$ stands for solar values.}

{As used in the previous works with the star-in-a-box setup \citep{kapyla2021star},
we use the enhanced luminosity method \citep{kapyla2020sensitivity} to reduce 
the gap between the dynamic and thermal timescales, which is infeasibly wide
otherwise. Therefore, our simulations have higher luminosity than in 
reality. The enhanced luminosity can be described by the luminosity ratio
$\Lratio = \mathcal{L}_{\rm sim}/\mathcal{L}_{\rm star}$ of the dimensionless 
luminosities of the simulation and the reference star. The dimensionless 
luminosity $\mathcal{L}$ is defined as
}
\begin{equation}
    \mathcal{L} = \sqrt{\frac{L}{G^3M^5/R^5}}.
\end{equation}
{With $\mathcal{L}_{\rm sim} = 4.8\times10^{-3}$ and 
$\mathcal{L}_{\rm star} = 1.3\times 10^{-10}$, 
$\Lratio = 3.6\times10^{7}$. This means that the gap between the dynamic and
thermal timescales is reduced by a factor of $\Lratio^{4/3}$. Further details
can be found in the Appendix of \cite{kapyla2020sensitivity}.
}
\subsubsection{Dimensionless units, initial and boundary conditions}
The boundary conditions are the same as those used by
\cite{kapyla2021star} and \cite{ortiz2023simulations}. The magnetic field is normal 
to the boundaries.
Such a condition is often thought to be more realistic than a vacuum boundary condition \citep{Yoshimura75}.
Impenetrable and stress-free conditions are used for the flow
such that the normal component is zero, and the normal derivatives of the 
tangential velocity components vanish. The temperature is symmetric across
the boundaries of the box, and the density satisfies a zero second-derivative condition.

For the units of length, time, density, and entropy we choose
\begin{eqnarray}
[x] = R,\ [t] = \sqrt{R^3/GM},\ [\rho] = \rho_0,\ [s] = c_{\rm P},
\end{eqnarray}
where $\rho_0$ is the initial density at the centre of the star. 
The magnetic fields are normalised by the equipartition field strength,
defined as $B_{\rm eq} = \langle \mu_0\rho\mUUU^2 \rangle^{1/2}$.

\subsubsection{Summary of 3D simulations}
We present five 3D MHD simulations at different
rotation rates. The simulations are characterised by dimensionless 
numbers, such as the Coriolis number ($\Co$), stellar Coriolis number
($\Costar$),
Taylor number ($\Ta$), the Reynolds number ($\Rey$),
magnetic Reynolds number ($\ReM$), the SGS ($\PraSGS$) and the
magnetic Prandtl number ($\PrM$), defined as
\begin{align*}
{\Co} &= \frac{2\Omega_0}{k_1 u_{\rm rms}^{\prime}}, 
& \Costar &= 2\Omega_0 \left( \frac {3MR^2}{L} \right)^{1/3},
& \Ta &= \frac{4\Omega_0^2 R^4}{\nu^2}, \\
\Rey &= \frac{u_{\rm rms}^{\prime}}{\nu k_1}, 
& \ReM &= \frac{u_{\rm rms}^{\prime}}{\eta k_1}, \\
  \PraSGS &= \frac{\nu}{\chi_{\rm SGS}},
& \PrM &= \frac{\nu}{\eta}, 
\end{align*}
where $k_1 = 2\pi/R$ and $u_{\rm rms}^{\prime}$ is the fluctuating velocity,
calculated as 
$u_{\rm rms}^{\prime} =
\left\langle
\left\langle \overline{u^2} \right\rangle_t -
\left[
\left(\left\langle \overline{u_r}\right\rangle_t\right)^2 +
\left(\left\langle \overline{u_\phi}\right\rangle_t\right)^2 +
\left(\left\langle \overline{u_z}\right\rangle_t\right)^2
\right]
\right\rangle_{\rm CZ}^{1/2}$,
{where the overbar denotes an azimuthal average, $\langle\cdot\rangle_t$ denotes 
a temporal average, and $\langle\cdot\rangle_{\rm CZ}$ denotes a spatial average over
the convection zone.} 
Direct solutions of the MHD equations capture the rotational influence on 
{the flow realistically. This is quantified by the Coriolis number, which
is the only parameter that can be correctly captured in 3D simulations \citep{kapyla2023simulations}.
}
{Other} parameters such as Prandtl and Reynolds numbers are either several 
orders of magnitude too large or too small \citep[e.g.][]{Kupka_Muthsam_2017_LRCA_3_1,Kapyla_et_al_2023_SSRv_219_58}.
The runs are listed in Table~\ref{tab:3Druns}{, where 
  the rms magnetic field in terms of the equipartition strength,
  the magnetic field strength close to the surface, 
  the rotation period,
  the amplitude of the relative differential rotation, and the
  dimensionless numbers described above are quoted for each run.}
\begin{table*}[t!]
\centering
    \begin{tabular}{ccccccccccc}
    \hline \hline
    Run  & $B_{\rm rms}/B_{\rm eq}$& $B_{\rm rms}^{\rm surf}\,[{\rm G}]$  & $\Co$& $\Costar$& $\Prot\,[\rm{d}]$& $\Delta\Omega/\Omega_0$& $\Ta$             & $\Rey$& $\ReM$ & Simulation time [yr]\\ \hline
    B1   &   0.43                  & 320  &  3.74 &  37.1    &  285    &      -1.59             & $4.44\times 10^7$ & 28.9  & 28.9     & 1365\\
    B2   &   0.42                  & 214  &  8.12 &  74.2    &  142    &      -0.81             & $1.78\times 10^8$ & 26.6  & 26.6     & 1415\\
    B5   &   0.45                  & 83.3 &  23.0 &  185     &  57     &      -0.32             & $1.11\times 10^9$ & 23.5  & 23.5     & 1244\\
    B10  &   2.24                  & 137  &  64.1 &  371     &  28     &      -0.11             & $4.44\times 10^9$ & 16.9  & 16.9     & 1241\\
    B20  &   2.42                  & 158  &  152  &  742     &  14     &      -0.04             & $1.78\times 10^{10}$ & 14.2  & 14.2  & 1324\\ \hline \hline
    \end{tabular}
    \caption{Summary of the 3D MHD simulations. The table lists the
    magnetic field strength normalised by equipartition, $B_{\rm rms}/B_{\rm eq}$, 
    {the magnetic field strength close to the surface $B_{\rm rms}^{\rm surf}$ in G,}
    the Coriolis numbers ${\rm Co}$ and ${\rm Co}_\star$, {the rotation period $\Prot$ in days,
    the amplitude of the relative differential rotation  
    at the equator $\Delta\Omega/\Omega_0$ ($\Delta\Omega=\Omega_{\rm top, eq}- \Omega_{\rm bot, eq}$, where $\Omega_0$ is the imposed rotation rate of the frame),}
    the Taylor number
    ${\rm Ta}$, the fluid and magnetic Reynolds numbers,
    ${\rm Re}$ and ${\rm Re_M}$, respectively. 
    {The last column shows the simulation time in years.} All runs have
    $\Pr_{\rm SGS}=1$ and $\Pr_{\rm M}=1${, and $288^3$ of grid resolution.}
    }
    \label{tab:3Druns}
\end{table*}
\subsection{Mean field model}
\label{sec:Mean-field-model}
The evolution of the mean magnetic field, here referring to
azimuthally averaged fields indicated by overbars, is given by
{the uncurled induction equation in the Weyl gauge \citep{brandenburg2005astrophysical}}
\begin{equation}
    \frac{\partial {\overline{\bm A}}} {\partial t} = \mUUU \times  \mBBB +  \mEMF - \eta \mu_0  \overline{\bm{J }},
\end{equation}
where $\mAAA$ and $\mBBB = \nabla \times \mAAA$ are the mean
magnetic vector potential and magnetic field, respectively. $\mEMF =
\alpha \mBBB - \eta_t \mu_0 \mJJJ$ is the mean
electromotive force, where $\alpha$ denotes the coefficient of the $\alpha$ effect and $\eta_t$ is the
turbulent magnetic diffusivity.

\begin{figure}[t!]\begin{center}
\includegraphics[width=\columnwidth]{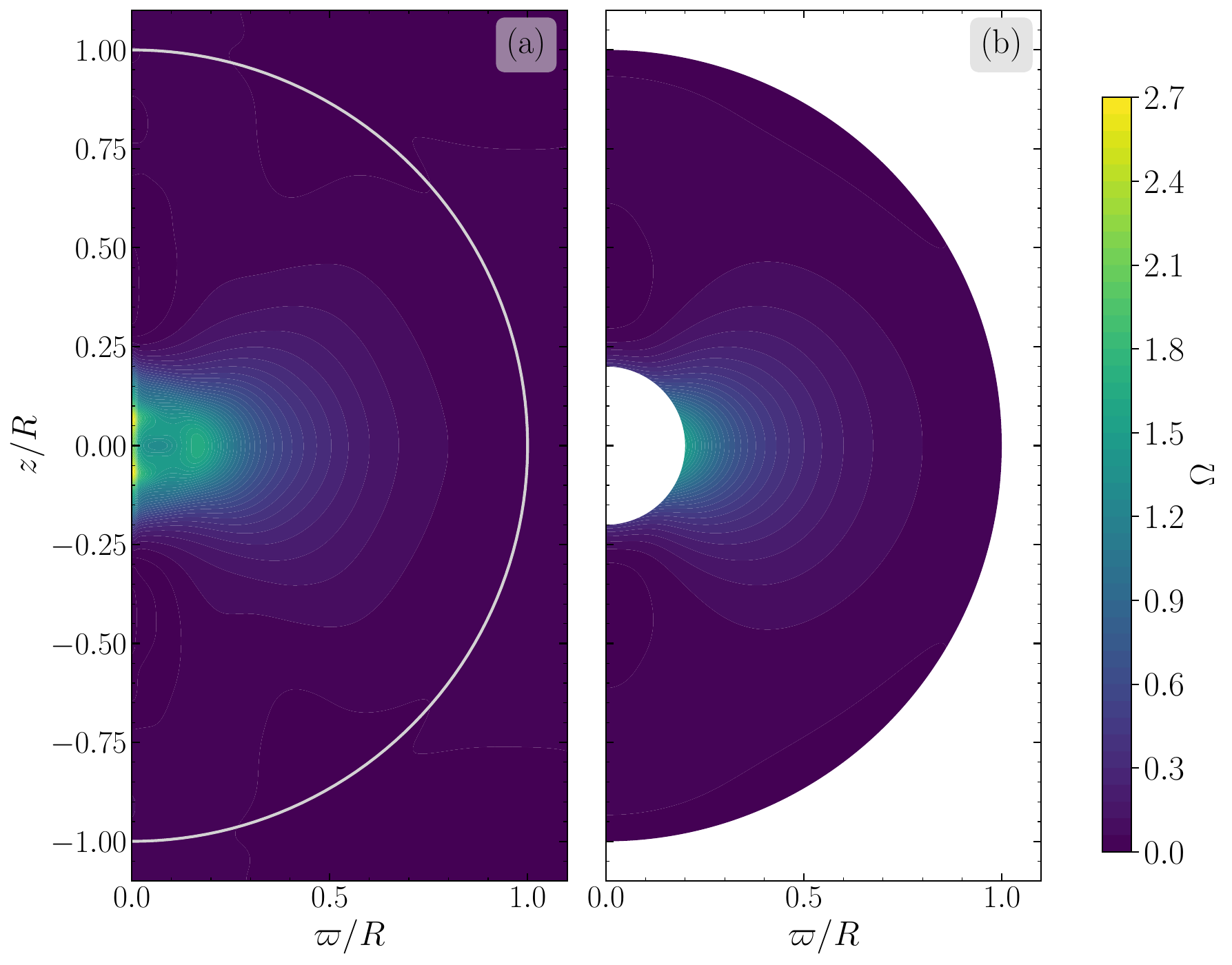}
\end{center}\caption[]{(a) Temporal and azimuthal average of the
  angular velocity profile of Run~B1. (b) Reconstructed angular
  velocity using the first three even-degree modes $\ell=0$, 2, and 4.
  The contours show $\Omega$ in spherical coordinates in the computational
  domain $0.2\leq r/R\leq1$.}
\label{fig:angular_vel}
\end{figure}

We estimate $\eta_t$ from the 3D MHD simulations, using
\begin{equation}
  \eta_t/\eta \sim \onethird \ReM.
\end{equation}
In Run B1,
$\ReM \sim 30$ and $\tilde{\eta} = 3 \times 10^{-4}$;
this yields $\tilde{\eta}_t \approx 3 \times 10^{-3}$, where 
the tilde denotes the normalised magnetic diffusivity,
$\tilde{\eta} = (RGM)^{-1/2} \eta$.
The prescribed mean flow is denoted by $\mUUU$. The angular velocity
was extracted from the 3D MHD run B1 by computing
azimuthal and temporal average of the azimuthal velocity component 
$U_\phi(\varpi,\theta)$, where $\varpi=r\sin\theta$. For comparison, one
of the mean-field runs used the rotation profile obtained from run B10.
The angular velocity in 
spherical coordinates is given by $\Omega(r,\theta) = U_\phi/(r \sin\theta)$.
The angular velocity was decomposed into Legendre polynomials,
 \begin{equation}
 	\Omega(r,\theta) = \sum_{\ell} \Omega_{\ell}(r) P_{\ell}\,(\cos \theta),
    \label{eq:omega}
 \end{equation}
and we retain the first three even-degree modes $\ell =0$, 2, and 4, which
capture the large-scale structure of the rotation profile.
Similar analyses were made previously by \cite{Rieutord+94} and 
\cite{Kapyla+11}.
In the present work, this decomposition is used as the input
rotation profile for the mean-field dynamo model. This allows us to isolate 
the dynamo action for a prescribed differential rotation profile. The left 
panel of Fig.~\ref{fig:angular_vel} shows the angular velocity from Run~B1, 
and the right panel shows its reconstruction from the first three even 
Legendre modes, as indicated by Eq.~\eqref{eq:omega}.
A list of all our mean-field runs presented in this work is given in
Appendix~\ref{sec:complete-set}.

Alternative polynomial expansions have been used in the literature to 
represent differential rotation in mean-field dynamo models. In particular, 
Gegenbauer polynomials have been employed in prescribed differential rotation 
parametrisation (e.g.\ \citealt{brandenburg2018strong}).
They have the advantage of converging more rapidly.
A comparison between the Legendre decomposition used the present work
and the corresponding Gegenbauer representation is shown in 
Appendix~\ref{sec:gegenbauer}.

For the mean-field models we use the {\sc Pencil
Code}\footnote{\href{https://github.com/pencil-code/pencil-code/}{github.com/pencil-code/}} in
spherical coordinates ($r,\theta,\phi$). The geometry of the model is defined 
by $0.2R  \leq r \leq R$, and $0 \leq \theta \leq \pi${, which corresponds
to the size of the convective envelope of our reference star and of the 3D model.}
We use the 3D simulations to constrain the inputs to the mean-field model and to define 
the physically relevant parameter space. The differential rotation
and estimates of the turbulent diffusivity are obtained from the 3D 
simulations{. The} $\alpha$ effect {is given by}
\begin{equation}
  \alpha(r,\theta) = \frac{1}{2} r\cos\theta\left[1-\operatorname{erf}\left(\frac{r-R}{\alpha_{\rm width}}\right)\right],
\end{equation}
{where $\alpha_{\rm width}=0.05$.}
{The $r\cos\theta$ factor represents the latitudinal dependence 
and reflects the antisymmetry of $\alpha$ about the equator.
Physically, this reflects the hemispheric antisymmetry of the kinetic helicity caused by interaction of rotation and stratification \citep{Steenbeck_et_al_1966_ZNatA_21_369}.
}
The strength of the $\alpha$ effect profile is varied
systematically to obtain similar growth rates to those found in 
the 3D runs.
This approach allows us to restrict the mean-field parameter 
space to values consistent with the 3D simulations. The resulting mean-field solutions 
are then evaluated by comparing their growth rates and large-scale magnetic field 
properties with those obtained in the 3D simulations.
Comparisons between 3D simulations and mean-field models are important 
in evaluating the validity of the mean-field approach \citep[e.g.][]{Kapyla_2025_LRSP_22_3}. 

\section{Results}
\subsection{Characterisation of the 3D simulations}
\begin{figure}[t!]\begin{center}
\includegraphics[width=\columnwidth]{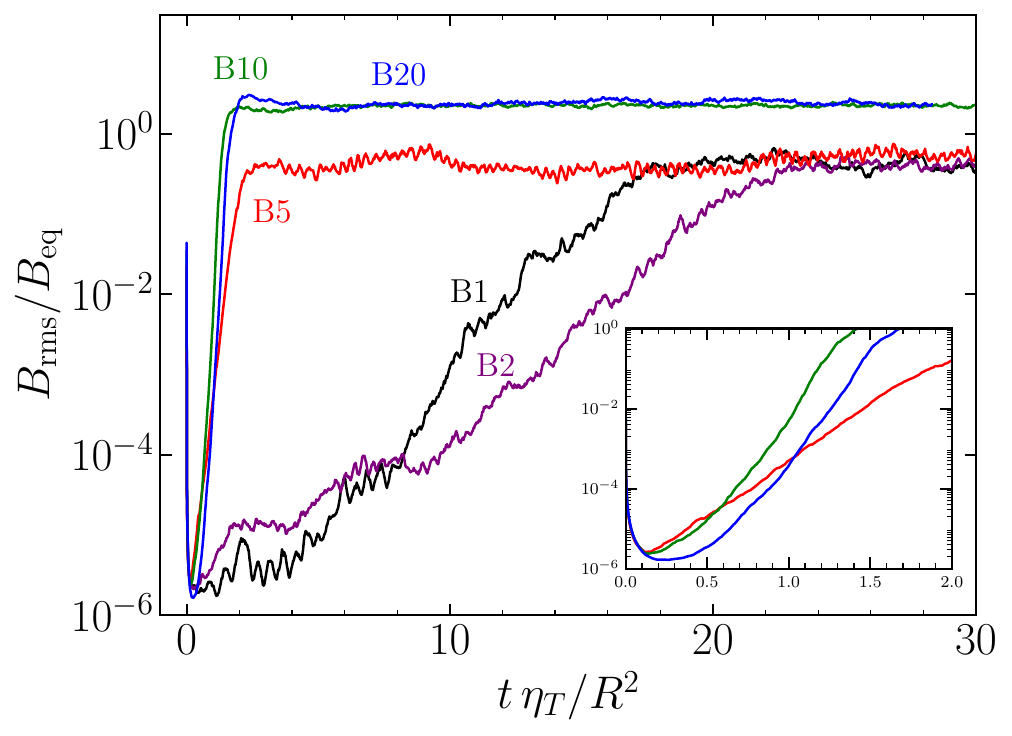}
\end{center}
\caption[]{Time evolution of the root-mean-square magnetic field, $B_{\rm rms}$ normalised by the equipartition field, $B_{\rm eq}$, for the 3D runs. 
The inset shows a zoom {$B_{\rm rms}$} for the higher rotation cases (B5, B10, and B20).}
\label{fig:brms_3d}
\end{figure}

Our goal is to compare the overall behaviour of mean-field models with
3D simulations. This includes the linear phase of the dynamo,
characterised by exponential growth, and the dynamo mode that is
realised in the saturated state.
Figure~\ref{fig:brms_3d} shows the evolution of the root-mean-square
magnetic field as a function of time for the five 3D runs.
The runs fall in roughly two categories in terms of the growth rate of
the magnetic field: the slow rotators (Runs~B1 and B2) take between 15
to 20 turbulent diffusion times ($\tau_{\rm diff}=R^2/\etaT$, where
$\etaT =\eta_t+\eta$) to
saturate whereas the more rapid rotators (Runs~B5, B10, and B20)
saturate already after one to two diffusion times. This suggests
a shift in the dominant dynamo mode.
{The effect of rotation is also seen in the amplification of the
magnetic field, as the saturation level of $B_{\rm rms}$ for the two most rapidly
rotating runs (B10 and B20) is above equipartition (see Table~\ref{tab:3Druns}).
}

\begin{figure}[t!]\begin{center}
\includegraphics[width=\columnwidth]{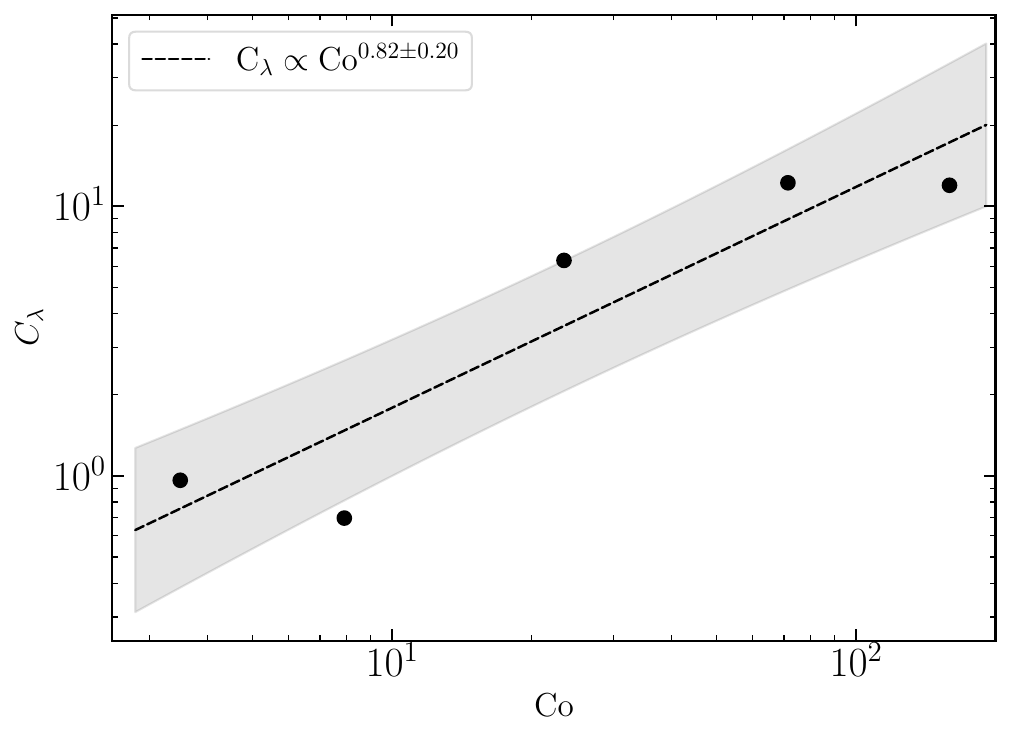}
\end{center}
\caption[]{$C_\lambda$ as {a} function of ${\rm Co}$ from the 3D runs.
The dashed line shows the best-fit power law 
{obtained from a least-squares fit in log-log space,} and the
shaded area shows the $\pm \sigma$ uncertainty of the fit.
The uncertainties in the individual $C_\lambda$ measurements are
smaller than the symbol size and are therefore not visible.
}
\label{fig:Clambda}
\end{figure}

\begin{figure}[t!]
\begin{center}
\includegraphics[width=\columnwidth]{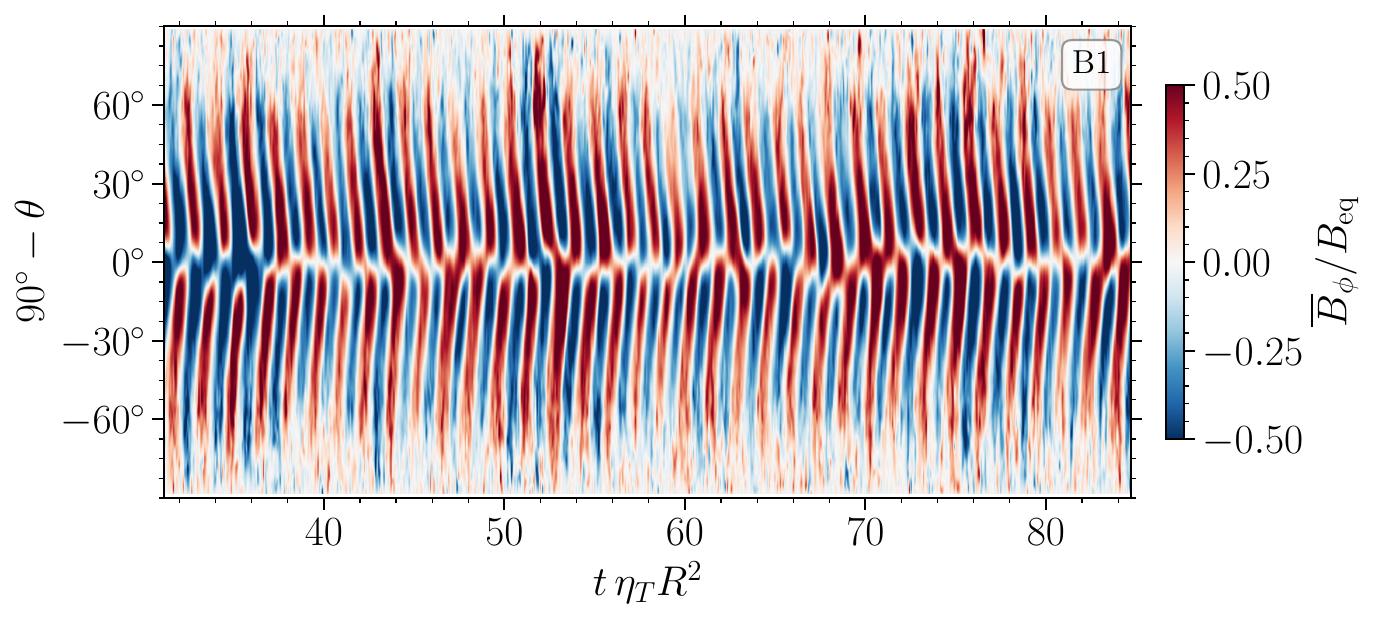}\\
\includegraphics[width=\columnwidth]{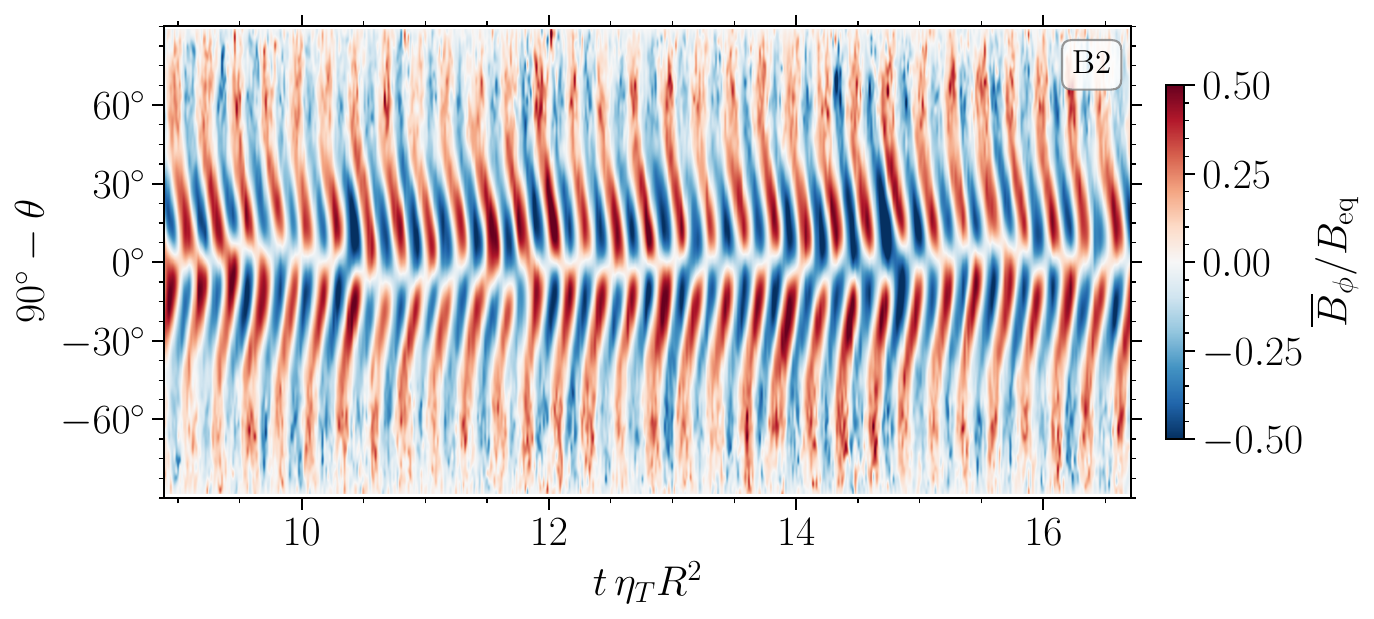}\\
\includegraphics[width=\columnwidth]{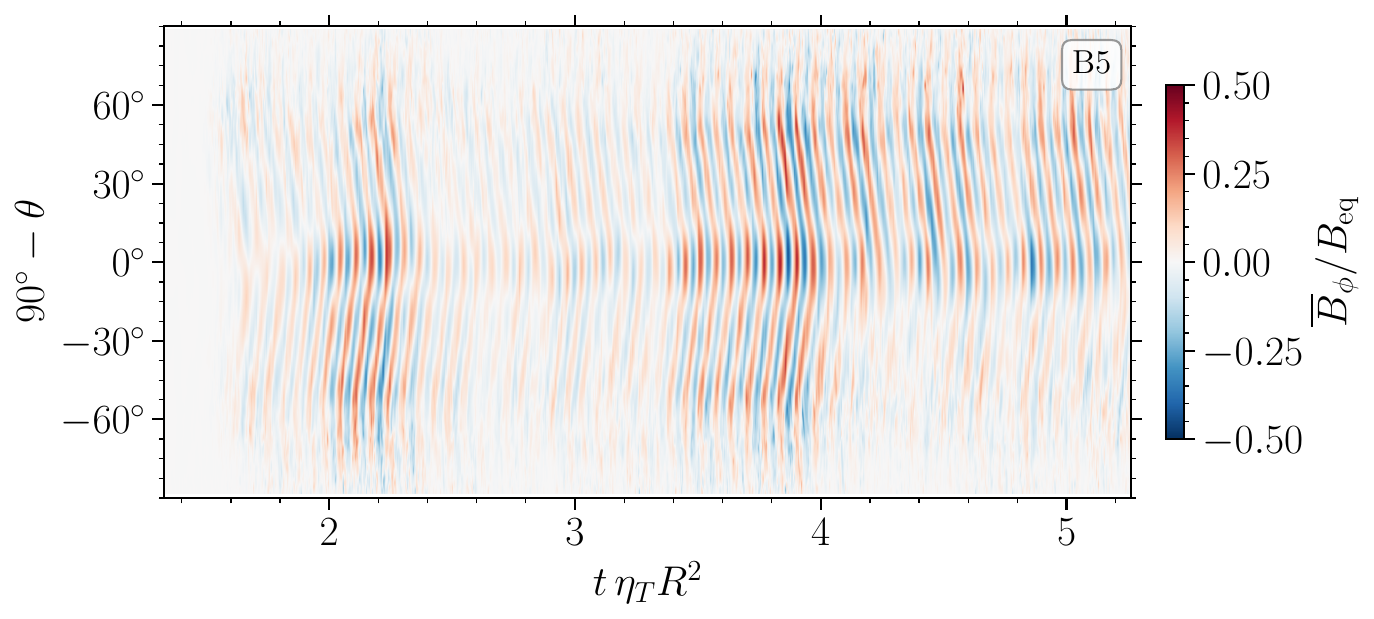}\\
\includegraphics[width=\columnwidth]{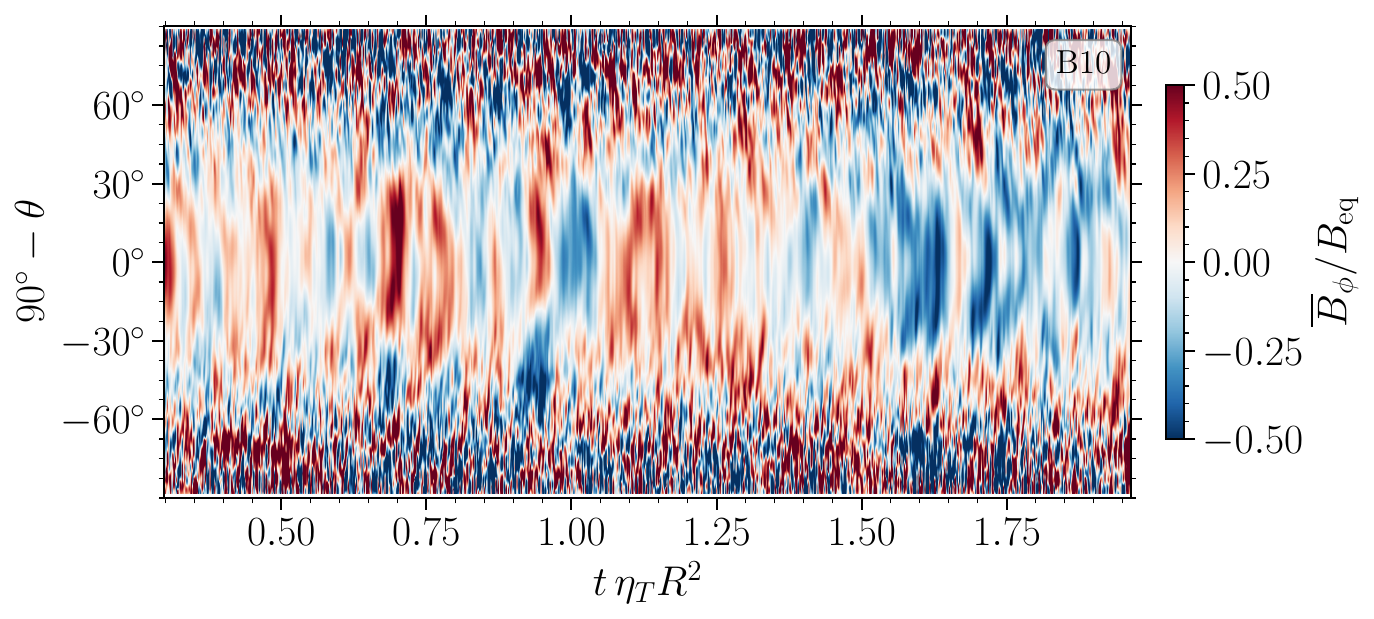}\\
\includegraphics[width=\columnwidth]{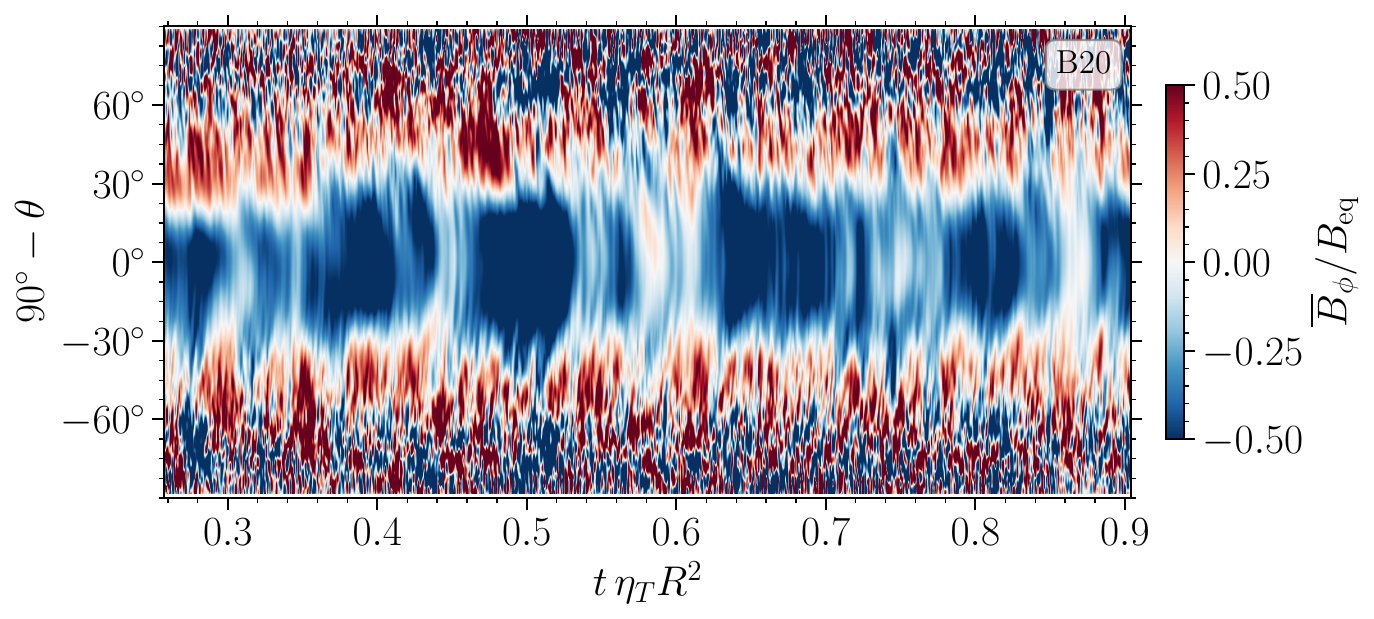}\\
\end{center}
\caption[]{Butterfly (time-latitude) diagrams of the mean magnetic field 
component $\overline{B}_{\phi}$ in units of the equipartition field $\Beq$
{for the 3D runs. }
}
\label{fig:butterfly3D}
\end{figure}

\begin{figure}[t!]\begin{center}
\includegraphics[width=\columnwidth]{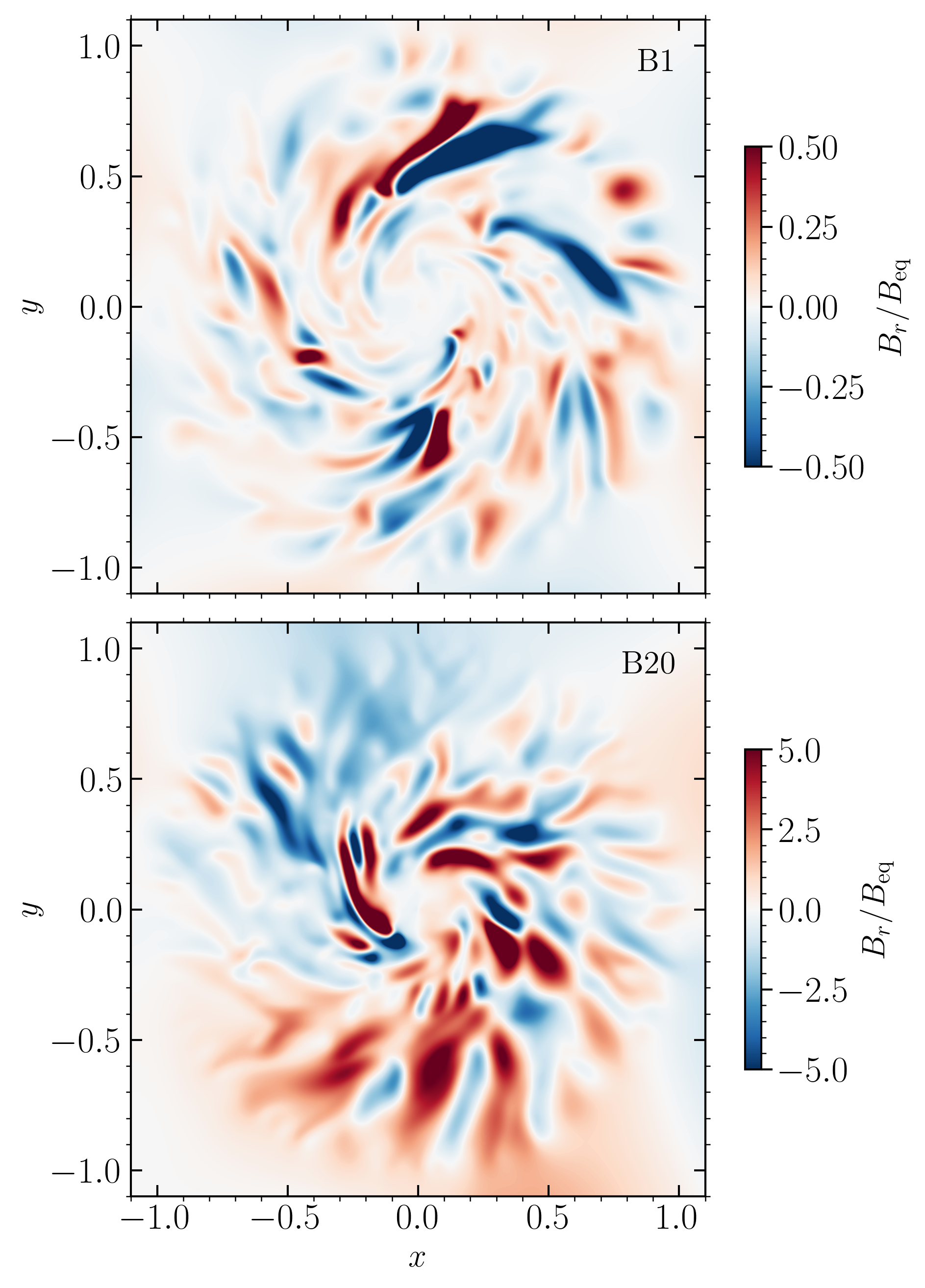}
\end{center}
\caption[]{Equatorial ($z=0$) view of the radial magnetic field $B_r$ 
of runs B1
and B20 from randomly chosen snapshots in the saturated dynamo regime.}
\label{fig:Br_eq}
\end{figure}

We characterise the global dynamo efficiency of the 3D MHD simulations
by measuring a non-dimensional growth rate of the root-mean-square
magnetic field, $C_\lambda$, via 
$B_{\rm rms} \propto \exp\left(\lambda t_{\rm diff}\right)$, where
$t_{\rm diff}$ is time measured in units of $\tau_{\rm
  diff}$, and therefore, $C_{\lambda} = \lambda t_{\rm diff}$.
Figure \ref{fig:Clambda} shows $C_{\lambda}$ as a function of the 
Coriolis number for the five 3D runs. 
The values of
$C_\lambda$ for each run are also listed in the second column of
Table~\ref{tab:dynamo_numbers}. The result shows a clear
dependence of the dynamo efficiency on rotation, with rapidly rotating
models having larger values of $C_{\lambda}$. This indicates the
importance of rotation in organising the flow for the growth of the
magnetic field.
However, despite the overall increase of $C_{\lambda}$ with rotation
rate, the growth rate does not follow a strictly monotonic trend, as
$C_\lambda(B2) < C_\lambda(B1)$, and $C_\lambda(B20)$ is slightly lower
than $C_\lambda(B10)$. 
Nevertheless, the data for $C_\lambda$ from the 3D simulations is
relatively well represented by a power law proportional to
$\Co^{0.82}$.
Figure~\ref{fig:butterfly3D} shows butterfly (time-latitude) diagrams
of the
azimuthally averaged toroidal component of the magnetic field, 
$\overline{B}_{\phi}(r,\theta,t)$. At slow and moderate rotation (Runs~B1,
B2, B5), the
simulations exhibit coherent axisymmetric
large-scale magnetic structures with regular polarity reversals, 
indicating a cyclic dynamo.
As the rotation increases (B10 and B20), the large-scale
organisation of the axisymmetric field becomes weaker. The
axisymmetric magnetic field shows small-scale
fluctuations and no clear cyclic behaviour. In addition, the field is more 
concentrated towards the poles.
Figure~\ref{fig:Br_eq} compares equatorial ($z=0$) cuts of the radial
magnetic field for runs B1 and B20. The latter shows a dominant 
large-scale non-axisymmetric magnetic structure whereas in the
former no single mode is immediately distinguishable.
This suggests a transition of the
dynamo mode at sufficiently rapid rotation such that a non-axisymmetric
$m=1$ mode becomes dominant. 
{A similar change from predominantly axisymmetric and often
cyclic dynamos to non-axisymmetric ones has been found at rapid enough
rotation in
}
simulations of partially {\citep[e.g.][]{Viviani_et_al_2018, navarrete2022origin}} and fully
convective main sequence stars \citep[e.g.][]{kapyla2021star}, {and is also expected
based on mean-field theory with an anisotropic $\alpha$ effect \citep{Ruediger80b,MB95}.}

\begin{table}[t!]
    \centering
    \begin{tabular}{c c c c c c c c}
        \hline \hline
        Run & $C_{\lambda}$ &
        $C_{{\alpha}_{\rm  sat}}$ &
        $C_{{\alpha}_{\rm  kin}}$ &
        $C_{\alpha_{\rm sat}}^{{\epsilon}_f }$ &  
        $C_{\alpha_{\rm kin}}^{{\epsilon}_f }$ &
        $-C_{\Omega_{\rm sat}}$ &
        $-C_{\Omega_{\rm kin}}$\\ \hline
        B1  & 0.96 & 1.28 & 1.39  & 1.51 & 1.63 & 6.63 & 6.36\\
        B2  & 0.70 & 1.47 & 1.55  & 1.70 & 1.77 & 7.43 & 7.34\\
        B5  & 6.30 & 2.11 & 2.20  & 2.18 & 2.27 & 9.56 & 8.08\\
        B10 & 12.2  & 2.40 & 3.03  & 2.60 & 3.19 & 14.1 & 5.40 \\ 
        B20 & 12.0  & 2.94 & 3.45  & 3.05 & 3.38 & 10.6 & 1.15 \\ \hline \hline
    \end{tabular}	-
    \caption{Summary of the non-dimensional growth rate and dynamo
      coefficients
    measured in the simulations. The table lists the values of 
    $C_{\lambda}$, {$C_{\alpha}^{\epsilon_f}$ (Eq.~\ref{eq:C_alpha_dispersion}), $C_\alpha$ (Eq.~\ref{eq:C_alpha_2D}), and $C_\Omega$ (Eq.~\ref{eq:C_Omega}).} {$C_{\alpha}^{\epsilon_f}$,} $C_\alpha$ and $C_\Omega$ are evaluated in both the kinematic and saturated regimes.} 
    \label{tab:dynamo_numbers}
\end{table}

\subsection{Mean-field characterisation of the 3D simulations}
\begin{figure}
    \centering
    \includegraphics[width=0.9\columnwidth]{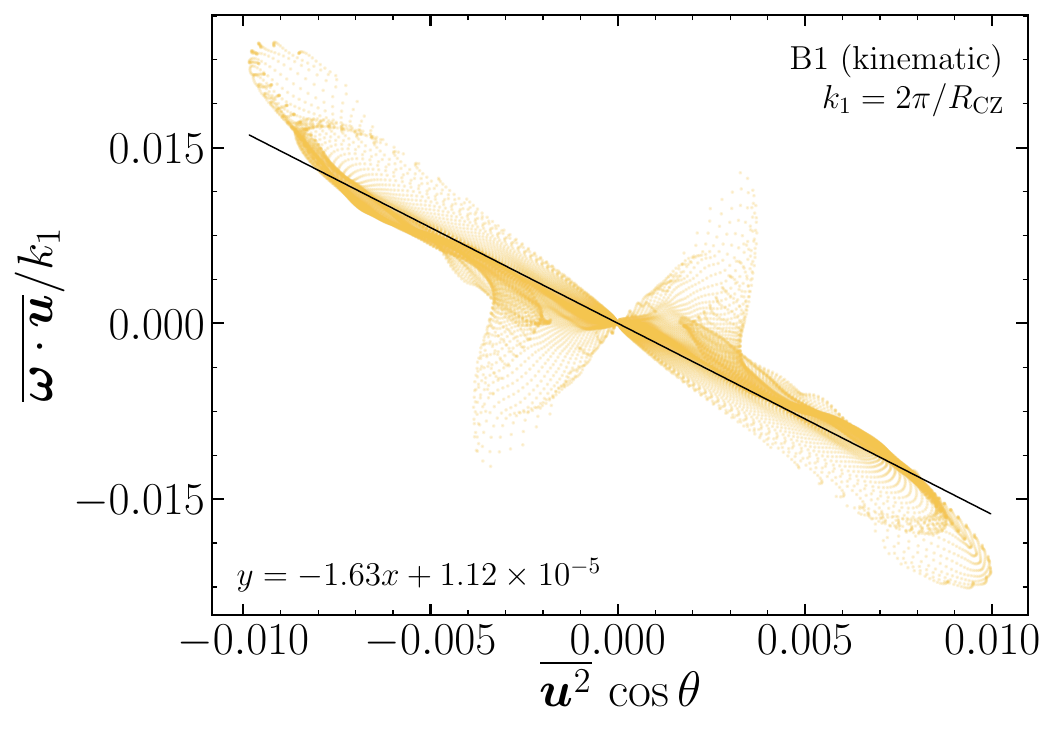} \\
    \includegraphics[width=0.9\columnwidth]{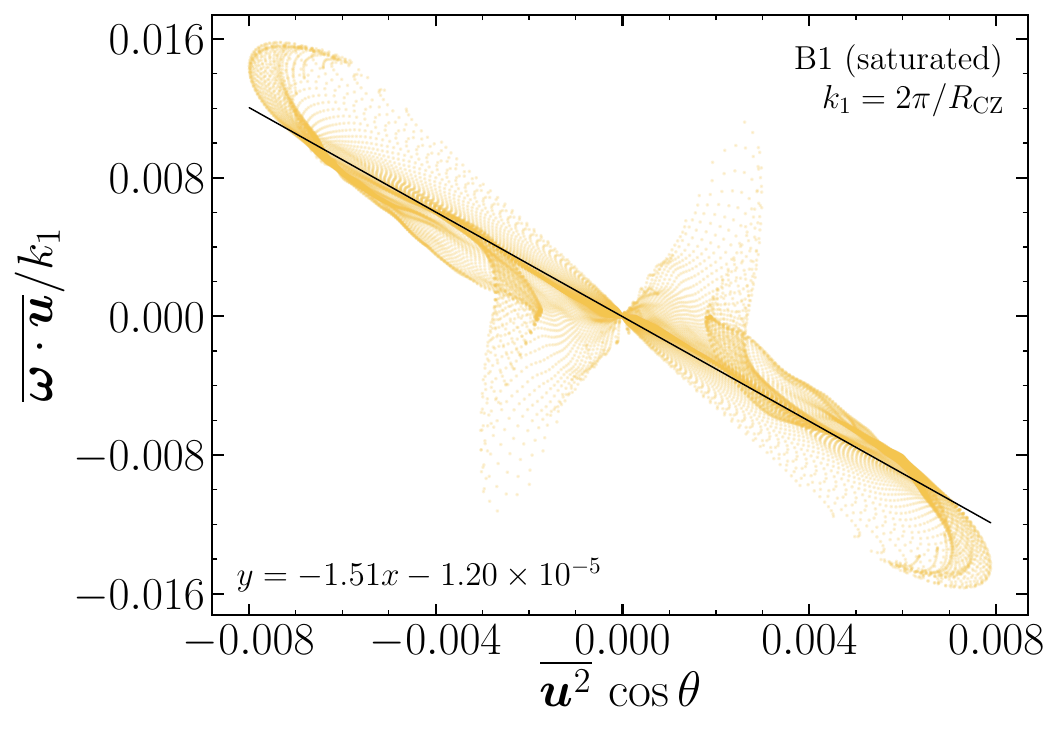}\\
    \caption{Scatter plot of the time-averaged kinetic helicity $\overline{\bm{\omega}\cdot \bm{u}}$ versus $\overline{\bm{u}^2}k_1\cos\theta$ for B1, 
    shown separately for the kinematic and saturated regimes. The solid line indicates a linear fit, whose slope defines $C_{\alpha}^{\epsilon_f}$. }
    \label{fig:C_alphaeps}
\end{figure}

\begin{figure*}
    \centering
    \includegraphics[width=1.\textwidth]{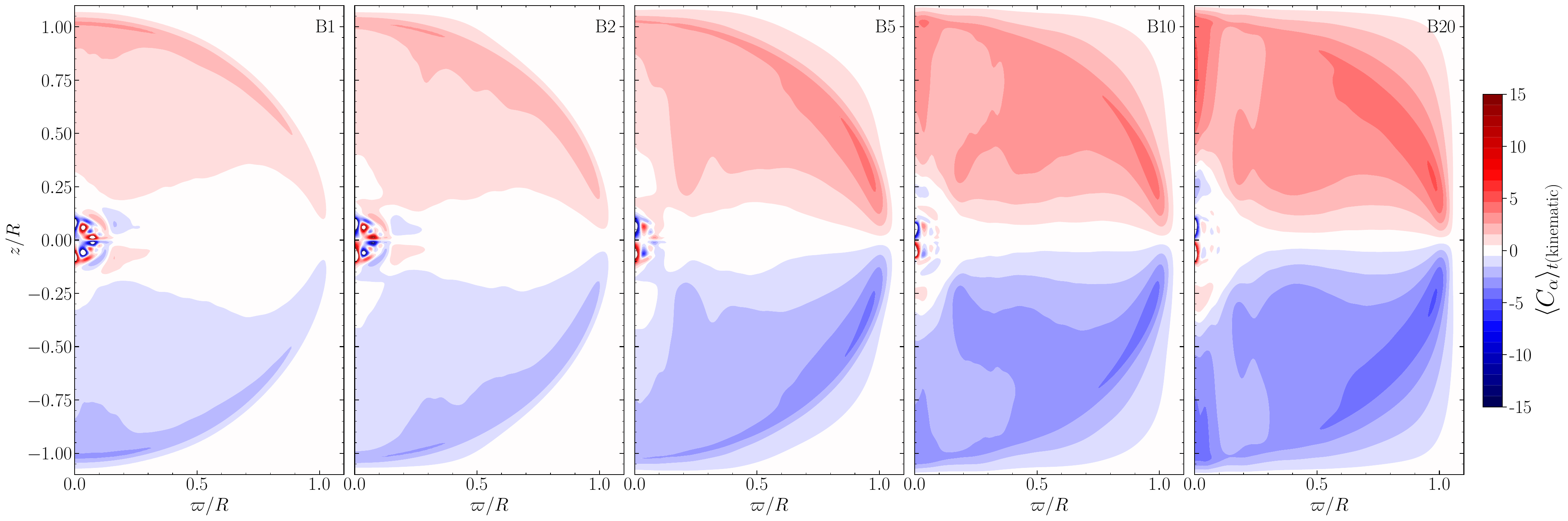} \\
    \includegraphics[width=1.\textwidth]{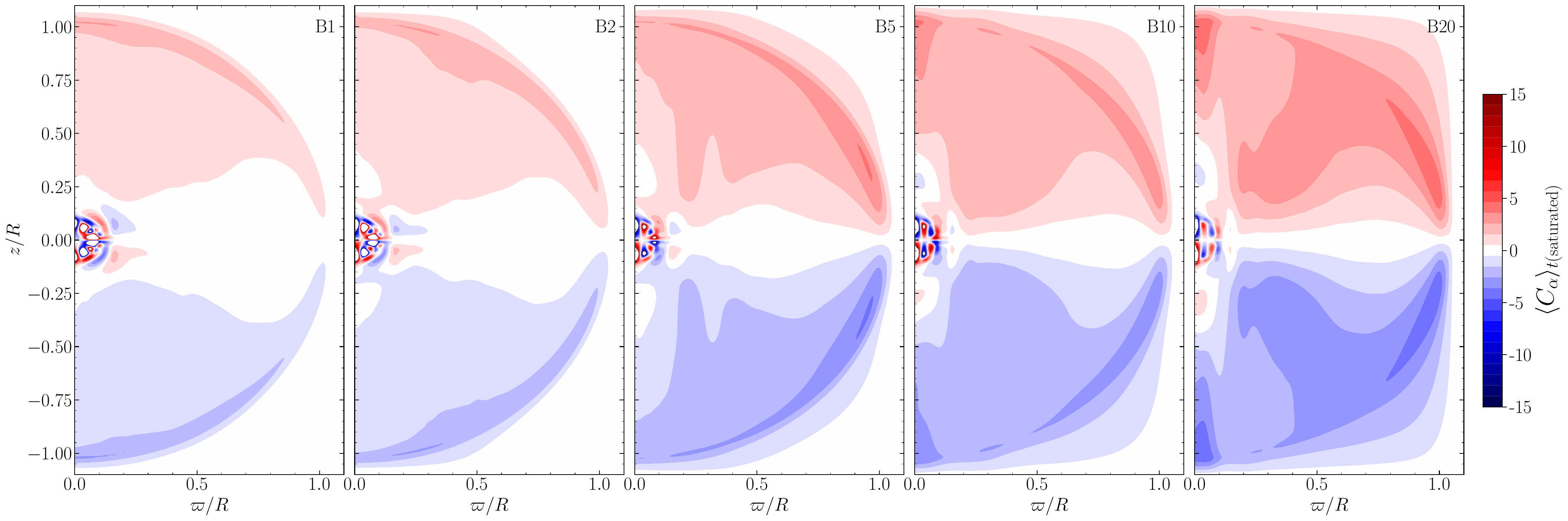}\\
    \caption{2D maps of $C_{\alpha}$ calculated with Eq.~\ref{eq:C_alpha} for the 3D runs.
The top panels correspond to the kinematic regime of the magnetic field, and the lower panels correspond to the saturated regime.}
    \label{fig:Calpha2d}
\end{figure*}

The $\alpha$ effect arises from helical turbulence and contributes 
to the generation of poloidal field from toroidal field \citep{Steenbeck_et_al_1966_ZNatA_21_369}. A first 
estimate of the $\alpha$ effect in the 3D simulations
can be obtained from a mean-field dispersion relation. We begin 
introducing the normalised kinetic helicity,
\begin{equation}
     \epsilon_f = \frac{\overline{\bm{\omega} \cdot \bm{u}}}{\overline{\bm{u}^2} k_f},
\end{equation}
which measures the degree of helicity relative to the kinetic 
energy at the dominant scale, $k_f$. In mean-field theory, the 
dynamo number associated with the $\alpha$ effect is defined as
$C_\alpha = \alpha/\eta_T k_1$. Under the estimates for isotropic
turbulence and first order smoothing approximation
\citep{brandenburg2005astrophysical}, the nondimensional parameter $C_\alpha=\alpha_0 R/\eta_\mathrm{t}$,
characterising the value of $\alpha_0$, can be approximated as \citep{Jabbari+14}
\begin{equation}\label{eq:C_alpha_dispersion}
C_{\alpha} \approx -\frac{\overline{\bm{\omega} \cdot \bm{u}}}{\overline{\bm{u}^2} k_1}
\approx -\epsilon_f \frac{k_f}{k_1},
\end{equation}
where $k_1$ (in the denominator of both expressions) is the characteristic large-scale wavenumber of the 
domain. To extract $C_{\alpha}$ directly from the simulations, we 
calculate the slope of the linear regression between 
${\overline{\bm{\omega} \cdot \bm{u}}}$ and ${\overline{\bm{u}^2}} k_1 \cos{\theta}$ 
evaluated within the convective zone.  The slope of this 
regression yields a 
measurement of $C_{\alpha}$, averaged separately over the 
kinematic and saturated phases of the magnetic field. We will 
refer to 
this estimation as $C_{\alpha}^{\epsilon_f}$. The resulting fits 
for run B1 during the kinematic regime are shown in 
Fig.~\ref{fig:C_alphaeps}. For completeness, these values and the 
values for the saturated regime are given in Table~\ref{tab:dynamo_numbers}.
While the dispersion relation gives us a single global estimate, we can 
also compute its local spatial structure. We evaluate the two-dimensional
distribution $C_{\alpha}(r,\theta)$ from the ratio of kinetic helicity to 
kinetic energy.
We also compare with Eq.~(18) of \cite{candelaresi_kinetic_2013}, who found
\begin{equation} \label{eq:C_alpha}
     C_{\alpha} = -\frac{ \overline{\bm{\omega} \cdot \bm{u}}}{\overline{\bm{u}^2} \left( 1 + 3{\rm Re_M}^{-1}\right)\,k_1}.
\end{equation}
Figure \ref{fig:Calpha2d} shows $C_{\alpha}(r,\theta)$ for the 3D runs.
The $\alpha$ effect exhibits an antisymmetric structure about the equator, with
opposite signs in the northern and southern hemispheres. The transition from 
the kinematic to the saturated regime is accompanied by a moderate reduction
in amplitude, consistent with the back reaction of the magnetic field on the flow.
To have a quantitative comparison with the dispersion definition, we 
extract a single value of $C_{\alpha}$ from these 2D maps. 
{We consider an average over the convective zone as follows :}
\begin{equation}\label{eq:C_alpha_2D}
    C_{\alpha} = \frac{\sum 2\cos\theta\,C_{\alpha}(r,\theta)}{N},
\end{equation}
{where the sum runs over the $N$ points in the meridional plane within the convective zone.}
We compute this quantity separately 
for the kinematic and saturated regimes.
$C_{\alpha}$ increases from B1 to B20 in both definitions, and is
consistently somewhat lower in the saturated regime than in the kinematic
one. The increase with rotation rate suggests that stronger rotation
enhances helical turbulence. This trend is also reflected in the
spatial structure of $C_\alpha$, with the faster rotating runs (B10
and B20) exhibiting broader and more intense regions near the poles
and at low latitudes compared to the
slowly rotating runs.

\begin{figure}[t!]\begin{center}
\includegraphics[width=\columnwidth]{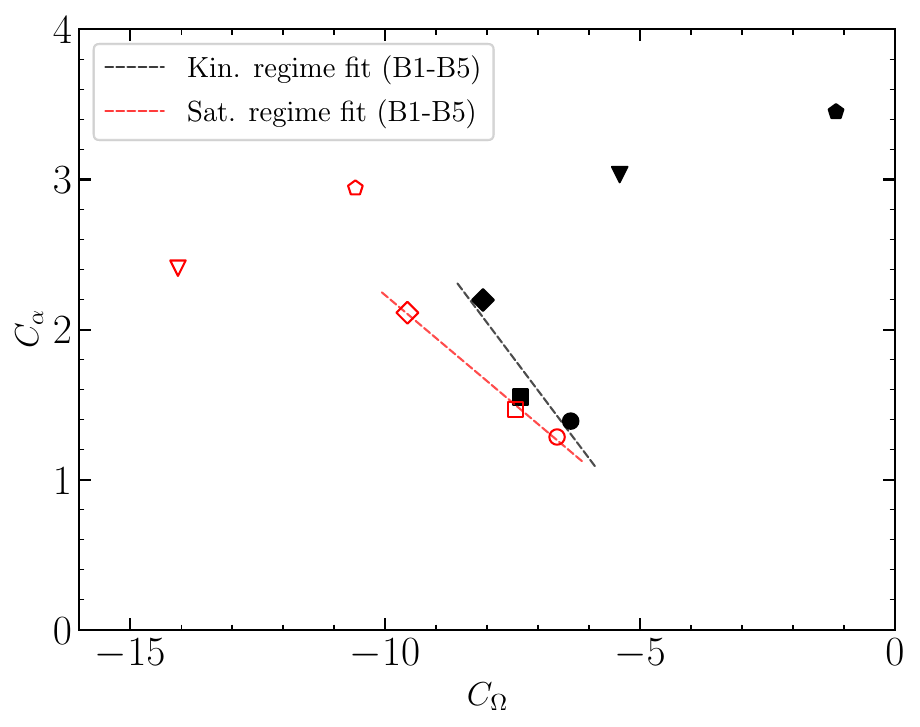}\\
\includegraphics[width=\columnwidth]{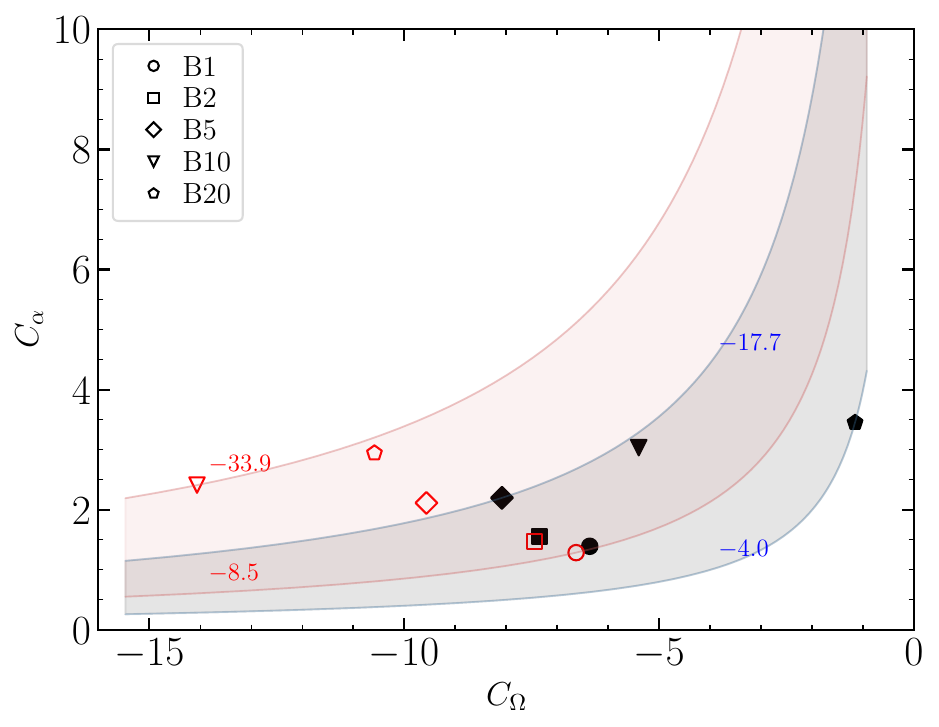}
\end{center}
\caption[]{Top: $C_\alpha$ as a function of $C_\Omega$ 
for the kinematic (filled black symbols) and saturated 
(open red symbols) regimes. 
Marker shapes correspond to different rotation rates: 
circles (B1), squares (B2), diamonds (B5), triangles 
(B10), and pentagons (B20). 
Dashed lines show linear fits to the low-rotation cases 
(B1–B5) in each regime, yielding 
$C_\alpha \approx (-0.45 \pm 0.20)\,C_\Omega$ and 
$C_\alpha \approx 
(-0.29 \pm 0.02)\,C_\Omega$ 
for the kinematic and saturated regimes, respectively. Bottom: The same diagram with shaded regions indicating the range 
of dynamo numbers $D=C_\alpha C_\Omega$ spanned by the simulations 
in the kinematic (grey) and saturated (light red) regimes across 
the runs ($D\approx -18$ to $- 4$ and $D\approx -34$ to $- 9$, 
respectively.)}
\label{fig:Raedler3D}
\end{figure}

Another important ingredient for large-scale dynamos is the $\Omega$ effect,
which quantifies the efficiency of differential rotation in generating toroidal field from poloidal field. It is defined as
\begin{equation}\label{eq:C_Omega}
	C_{\Omega} = \frac{\Delta \Omega}{\eta_T k_1^2}.
\end{equation}
where $\Delta \Omega$ is the estimated angular velocity difference from the radial
{shear defined as the difference between the equatorial angular
velocity near the surface ($\Omega_{\rm top, eq}$) and at the base of 
the convective zone ($\Omega_{\rm bot, eq}$), that is 
$\Delta\Omega=\Omega_{\rm top, eq}- \Omega_{\rm bot, eq}$.
}
The values for $C_\Omega$ are calculated for the kinematic and saturated regimes separately,
and they are given in the last two columns in Table~\ref{tab:dynamo_numbers}.
The top panel of Fig.~\ref{fig:Raedler3D} shows $C_{\alpha}$ as a function of  $C_\Omega$.
Here, $C_{\alpha}$ is obtained from the weighted average of the
two-dimensional maps using Eq.~\eqref{eq:C_alpha_2D}.
In the kinematic regime, the magnitude of $C_{\Omega}$ generally increases with rotation rate, 
although this trend becomes less regular at the highest rotation rates, B10 and B20. In the saturated regime, 
the magnitude of $C_{\Omega}$ increases from B1 to B10, while B20 departs from this behaviour, suggesting
a modification of the dynamo properties at rapid rotation.
{To quantify this trend, linear fits to the data of the slowly rotating cases 
(B1-B5) result in $C_\alpha \approx (-0.45 \pm 0.20)\,C_\Omega$ for the kinematic 
regime and $C_\alpha \approx (-0.29 \pm 0.02)\,C_\Omega$ 
for the saturated regime.}
At the same time, $C_{\alpha}$ is systematically lower in the saturated regime compared to the 
kinematic one. This reduction is consistent with $\alpha$ quenching, arising from the 
backreaction of the Lorentz force on the flow.
The bottom panel of Fig.~\ref{fig:Raedler3D} shows an approximate R\"adler diagram-like representation with curves of constant the dynamo 
number, $D=C_\alpha C_\Omega$.
An actual R\"adler diagram only shows the marginally excited states \citep{BLDSK25}.
The simulations occupy a band of negative dynamo numbers, spanning 
approximately from $D\approx-18$ to $-4$ in the kinematic regime and from $D\approx-34$ to $-9$ in the saturated 
regime. 

\subsection{Mean-field parameter study}
\label{sec:Mean-field-parameter-study}
Across the different runs, we varied the strength of the $\alpha$ effect, 
by varying the value of $C_\alpha$ from $0$ to $2.97$.
We mainly used the 
rotation profile of the 3D run B1, and also considered scaled versions of this profile
using different scale factors. In addition to $\alpha^2\Omega$
dynamos, we explored $\alpha^2$ models and runs in which we used the
$\alpha\Omega$ approximation, where
{the $\alpha$ tensor is restricted to its $\phi\phi$ component. Therefore,
the $\alpha$ effect contributes only to the poloidal field generation \citep{brandenburg2018strong}.
}
Table~\ref{tab:mf_reduced} summarises our representative mean-field runs, 
while the complete set of simulations is given in Appendix~\ref{sec:complete-set}.

\begin{figure}[t!]\begin{center}
\includegraphics[width=\columnwidth]{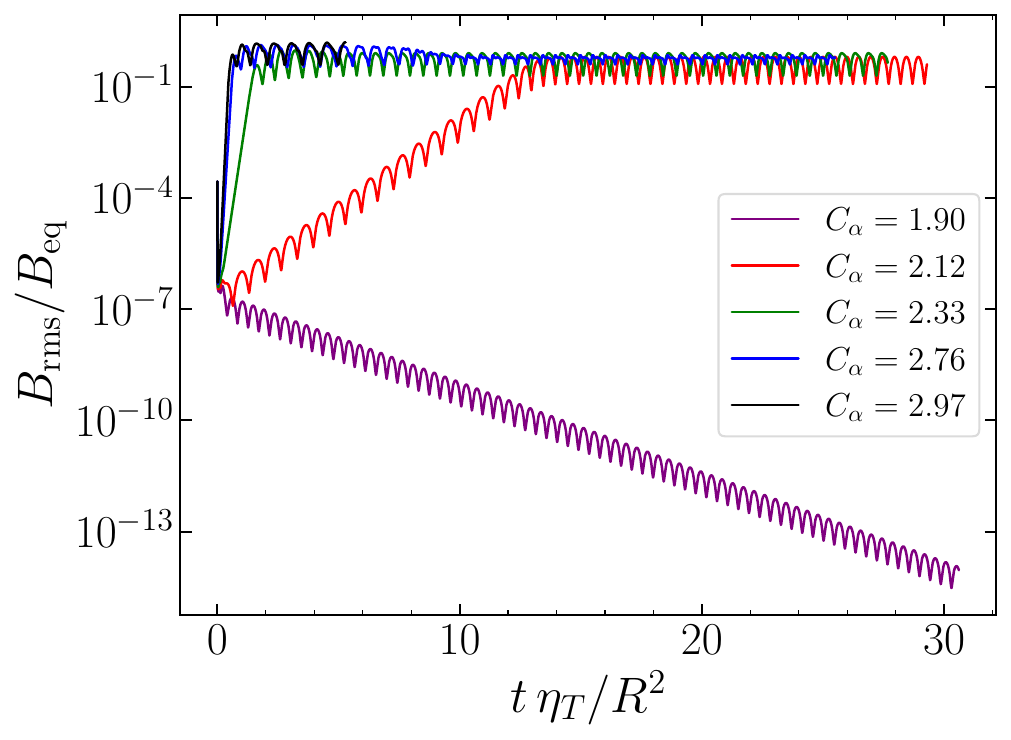}
\end{center}
\caption[]{Time evolution of $B_{\rm rms}$ in mean-field runs with fixed 
rotation rate ($\Omega_{B1}$) and varying $C_\alpha$, as indicated by the legend.}
\label{fig:brms_mf}
\end{figure}

For the subset of runs with fixed rotation rate and without enforcing the 
$\alpha\Omega$ approximation, we use the rotation profile of the 3D run B1, and 
varied the strength of the $\alpha$ effect. The run ${\rm A_{B1}}$ (with $C_{\alpha}=1.90$)
is subcritical, with no growth of the magnetic field, 
while the runs with $C_\alpha \geq 2.12 $  are supercritical
(the marginal value is around $2.0$).
Figure~\ref{fig:brms_mf} shows the time evolution of $B_{\rm rms}$ of these runs.
The $\alpha^2\Omega$ models exhibit growing $B_{\rm rms}$ even in the
$\alpha^2$ limit, that is in the absence of differential
rotation. This shows that the
$\alpha$ effect alone can sustain dynamo action within our model. When the
rotation profile of B1 ($\Omega_{\rm B1}$) and its scaled version 
{($2\Omega_{\rm B1}$)} are included in the runs with $C_\alpha=2.97$, 
the growth rate increases, indicating that shear enhances the dynamo, but it is not
required for its operation.

\begin{figure}[t!]\begin{center}
\includegraphics[width=\columnwidth]{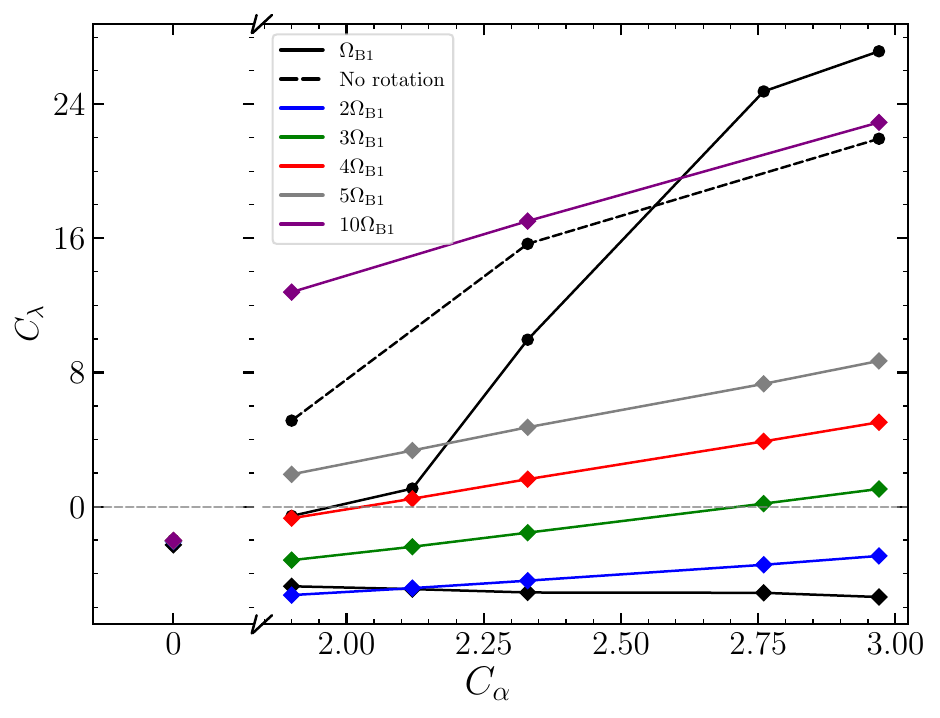}
\end{center}
\caption[]{
Dimensionless growth rate $C_\lambda$ as function $C_\alpha$. Each colour represents different rotation
rate. The diamonds are for the runs with the $\alpha\Omega$ approximation, and the circles are
for the runs without it. }
\label{fig:lambda_alpha}
\end{figure}

We estimate $C_{\alpha,{\rm crit}}$, which denotes the critical value for 
which the mean field runs become marginally supercritical. 
This is calculated from the zero crossings ($C_\lambda=0$) in 
Fig.~\ref{fig:lambda_alpha} using linear interpolation. 
We find that for the $\alpha^2\Omega$ model with $\Omega_{\rm B1}$,
the threshold lies at $C_\alpha\approx2.0$.
In contrast, runs with the $\alpha\Omega$ approximation and moderate shear 
($\Omega_{\rm B1}$ and $2\Omega_{\rm B1}$) are subcritical in the 
parameter range explored. This indicates that the imposed shear is not strong
enough to produce the toroidal field from the poloidal field. For stronger shear,
the simulations become supercritical. For each explored value of 
$\alpha$, there is a minimum shear needed for the $\alpha\Omega$ dynamo to operate.
For completeness, we also present the
dependence of $C_{\lambda}$ on the dynamo
number $C_{\alpha}C_{\Omega}$ in
{Appendix}~\ref{Growth-rate-vs-dynamo-number}.
\begin{table}
\centering
\begin{tabular}{c c c c}
\hline \hline
Run & $\Omega$ & $C_\alpha$ & $C_\lambda$ \\
\hline
\multicolumn{4}{c}{{$C_\alpha$ at fixed rotation}} \\
\hline
${\rm{A_{\rm B1} (\alpha \Omega)}}$ & $\Omega_{\rm B1}$ & $1.90 $ & $-4.79$ \\
${\rm{B_{\rm B1} (\alpha \Omega)}}$ & $\Omega_{\rm B1}$ & $2.12$ & $-4.92$  \\
${\rm{C_{\rm B1} (\alpha \Omega)}}$ & $\Omega_{\rm B1}$ & $2.33$ & $-5.12$  \\
${\rm{D_{\rm B1} (\alpha \Omega)}}$ & $\Omega_{\rm B1}$ & $2.76$ & $-5.15$  \\
${\rm{E_{\rm B1} (\alpha \Omega)}}$ & $\Omega_{\rm B1}$ & $2.97$ & $-5.40$  \\
\hline
\multicolumn{4}{c}{{Dependence on rotation (full model)}} \\
\hline
${\rm{E_0}}$ & $\Omega_0$ & $2.97$ & $21.94$  \\
${\rm{E_{\rm B1}}}$ & $\Omega_{\rm B1}$ & $2.97$ & $27.15 $  \\
${\rm{E_{{\rm B1} \times 2}}}$ & $2\Omega_{\rm B1}$ & $2.97$ & $30.21$ \\
\hline
\multicolumn{4}{c}{{$\alpha\Omega$ comparison}} \\
\hline
${\rm{E_{0} (\alpha\Omega)}}$   & $\Omega_0$ & $2.97$ & $-2.02$ \\
${\rm{E_{B1} ({\alpha\Omega})}}$& $\Omega_{\rm B1}$ & $2.97$ & $-5.40$ \\
${\rm{E_{B1\times2}({\alpha\Omega})}}$ & $2\Omega_{\rm B1}$ & $2.97$ & $-2.94$ \\
\hline \hline
\end{tabular}
\caption{Representative mean-field runs illustrating the dependence 
of the dimensionless growth rate $C_\lambda$ on $\alpha$ effect and rotation. 
The upper block shows the variation with $C_\alpha$ at fixed rotation 
$\Omega_{\rm B1}$. The middle block isolates the effect of rotation for 
$C_\alpha = 2.97$. The lower block compares the full model 
with runs where the $\alpha\Omega$ approximation is enforced.}
\label{tab:mf_reduced}
\end{table}

\begin{figure}
    \centering
    \includegraphics[width=\columnwidth]{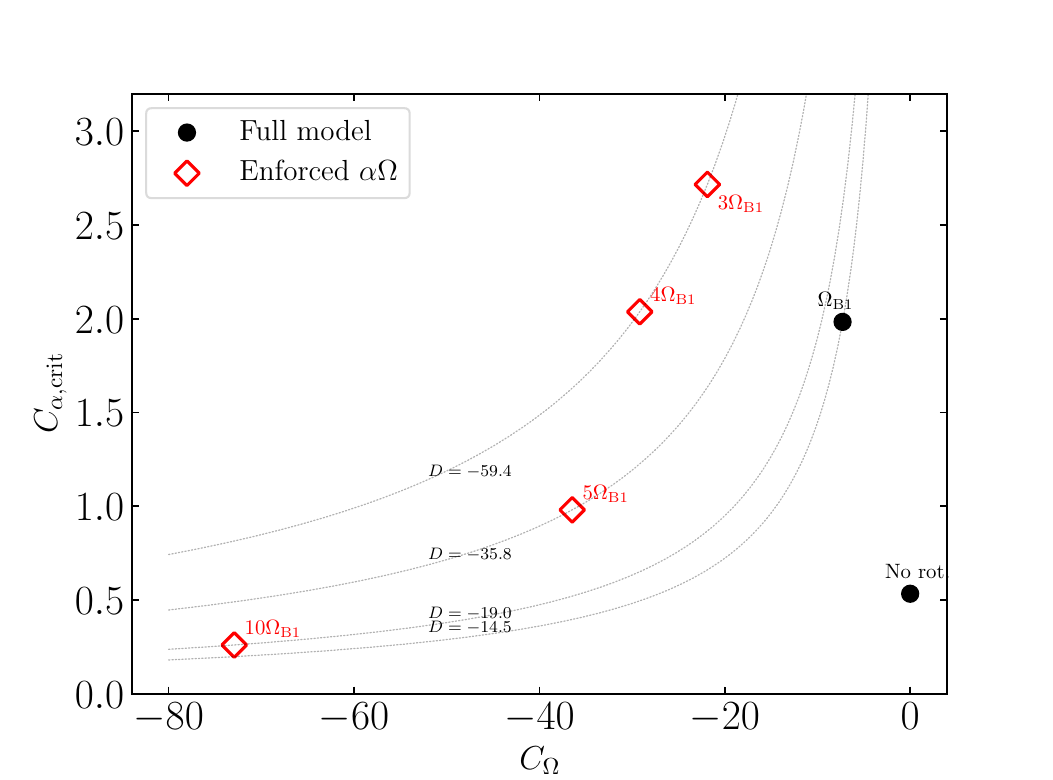}
    \caption{R\"adler diagram for the mean-field models. The critical values of 
    $C_{\alpha,\text{crit}}$ required for dynamo excitation are shown as function 
    of $C_\Omega$. Filled black circles correspond to the full mean-field dynamos
    model, while open red diamonds indicate simulations with $\alpha\Omega$ 
    approximation. The dotted curves represent lines of constant dynamo number 
    $D=C_\alpha C_\alpha$.}
    \label{fig:raedler_mf}
\end{figure}

Figure~\ref{fig:raedler_mf} 
shows the corresponding R\"adler diagram, where $C_{\alpha,{\rm crit}}$ is plotted 
as a function of $C_\Omega$. The simulations shown here have a negative dynamo number $D$,
and lines of constant $D$ are indicated. The runs with enforced $\alpha\Omega$ using the 
rotation profiles $3\Omega_{\rm B1}$ and $4\Omega_{\rm B1}$ share a similar dynamo number,
$D\approx -59$, suggesting that they are in a similar dynamo regime. 
The runs using the rotation profile $5\Omega_{\rm B1}$, $10\Omega_{\rm B1}$, and $\Omega_{\rm B1}$, have $D\approx-36$, 
$-19$, and $-14$, respectively.

\subsection{Role of differential rotation on the magnetic decay}
In the early days of dynamo theory, when the physical reality of the
$\alpha$ effect was not yet universally accepted, \cite{Chandrasekhar56}
proposed to explain the Earth's magnetic field as the result of a strongly
reduced decay rate due to the presence of differential rotation; see
\cite{Brandenburg11} for a discussion of the historical context of
his paper.
This proposal seemed surprising, but it is indeed possible that certain
flow geometries can significantly slow down the magnetic decay.
This can be the case when the mean-field evolution depends on the magnetic
field at all earlier times, which is known as the memory effect \citep{HB09}.
In this context, we mention here that a particular one-dimensional
irrotational flow with such properties was studied by \cite{Raedler+11},
where the actual turbulent decay rate of a passive scalar was found to be 200 times slower
than the naive estimate of the decay rate that ignores the memory effect.

\begin{figure}[t!]\begin{center}
\includegraphics[width=\columnwidth]{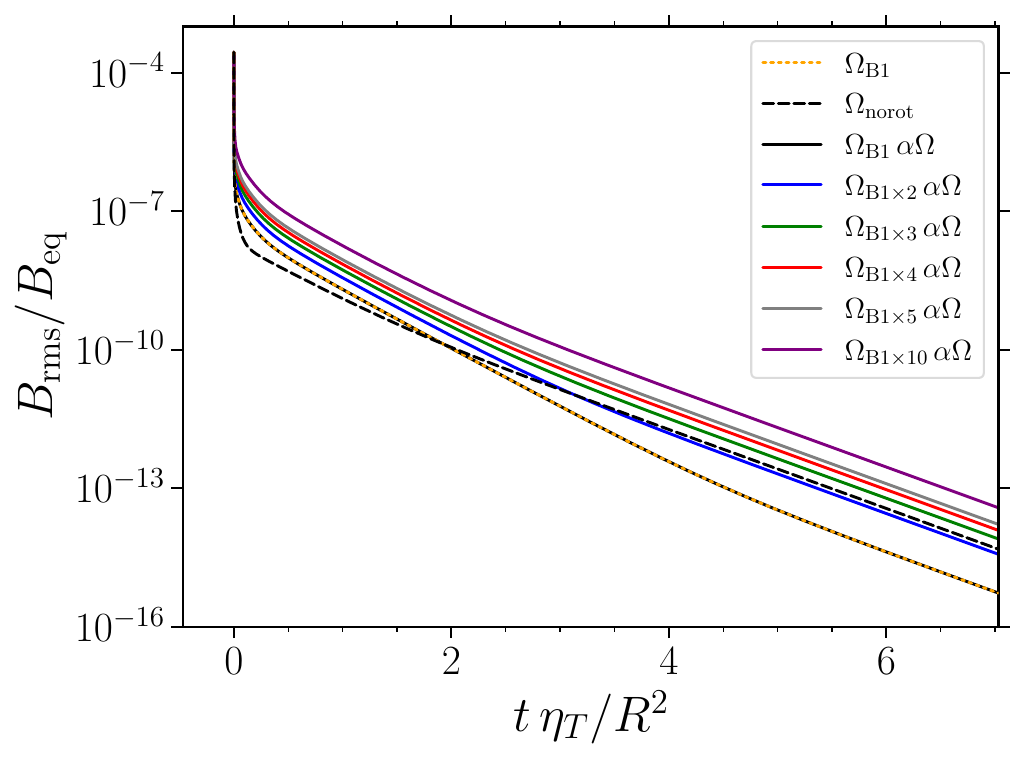}
\end{center}
\caption[]{Time evolution of $B_{\rm rms}$ of mean-field runs with 
{$C_\alpha=0$ and varying strength of the differential rotation.}
The dashed line marks the case without differential rotation.
}\label{fig:brms_mf_alpha0}
\end{figure}

In view of earlier ideas that the magnetic fields of red giants could
be fossil ones, we now compute the decay rates of our models in the
absence of an $\alpha$ effect.
The result is shown in Fig.~\ref{fig:brms_mf_alpha0}, where we show
the decay of $B_\mathrm{rms}$ versus time for different strengths of
differential rotation.
It turns out that differential rotation always speeds up the decay of
the magnetic field.
The weakest differential rotation leads to the fastest
decay, while the run with ten times larger differential rotation decays more slowly,
but the non-rotating case decays more slowly still.
Furthermore, unlike the case studied by \cite{Chandrasekhar56}, which
applied to a non-turbulent medium, we consider here a turbulent one
where the decay rate in the absence of motions is already based on a
turbulent magnetic diffusivity.
Within the framework of the present study, the flows expected to exist in red
giants are unlikely to prolong the decay rates, making a fossil magnetic
field hypothesis {unlikely to explain the observed surface fields 
and indicating that they} are indeed dynamo-generated ones.

\subsection{Modes of dynamo action in mean-field models}
\begin{figure*}[t!]
    \centering
    \includegraphics[width=0.33\linewidth]{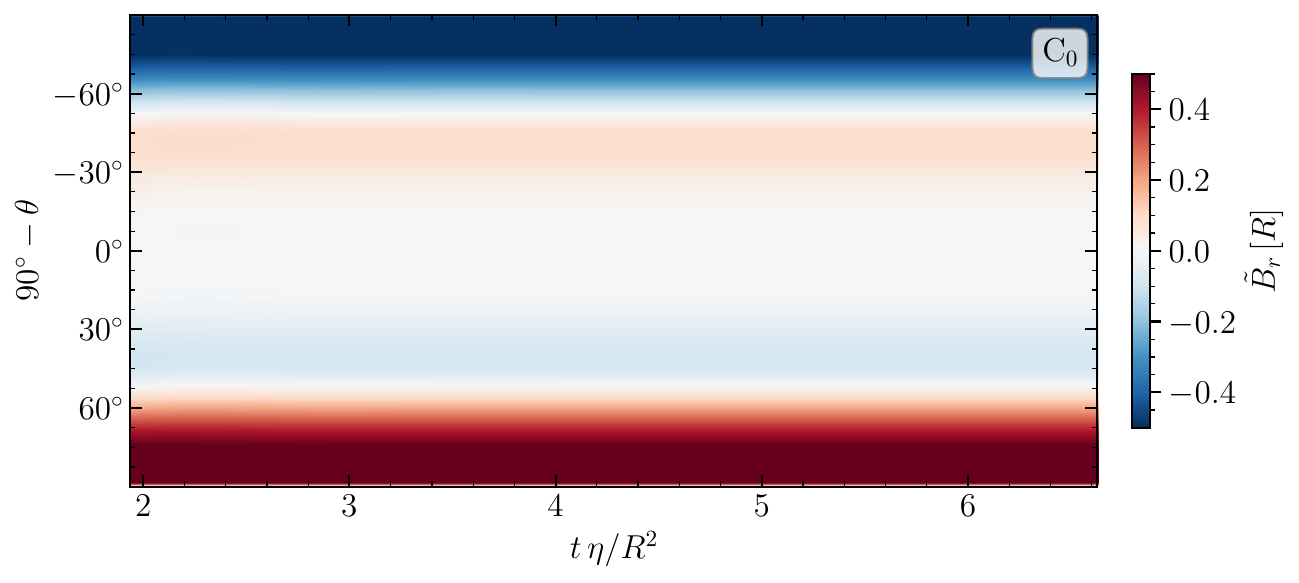}
    \includegraphics[width=0.33\linewidth]{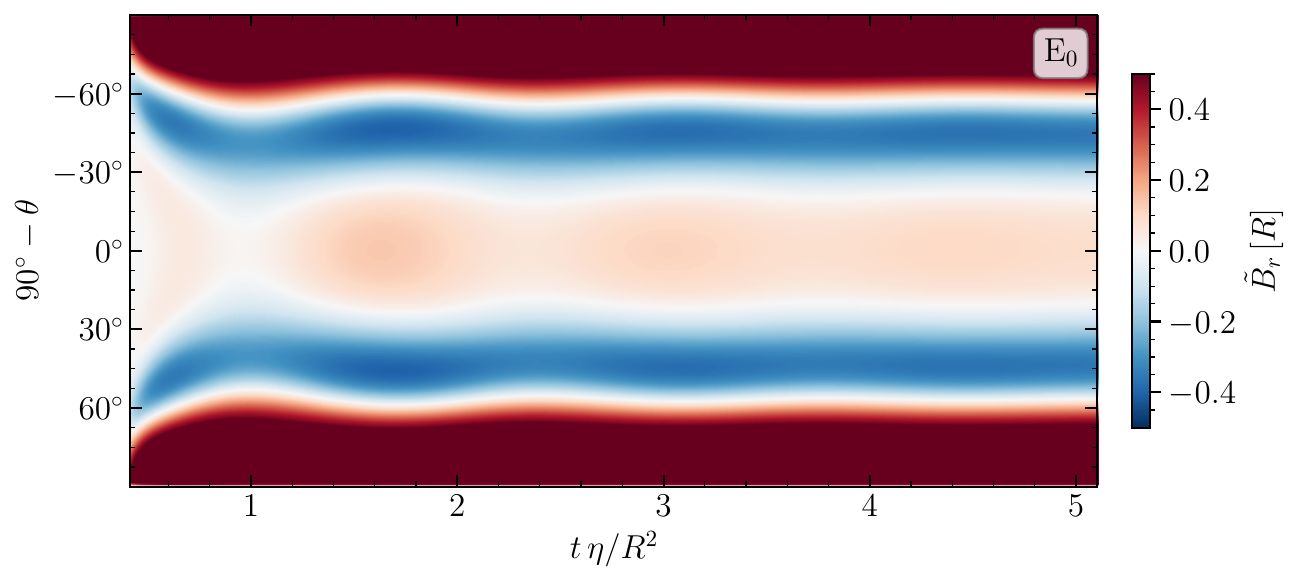}
    \includegraphics[width=0.33\linewidth]{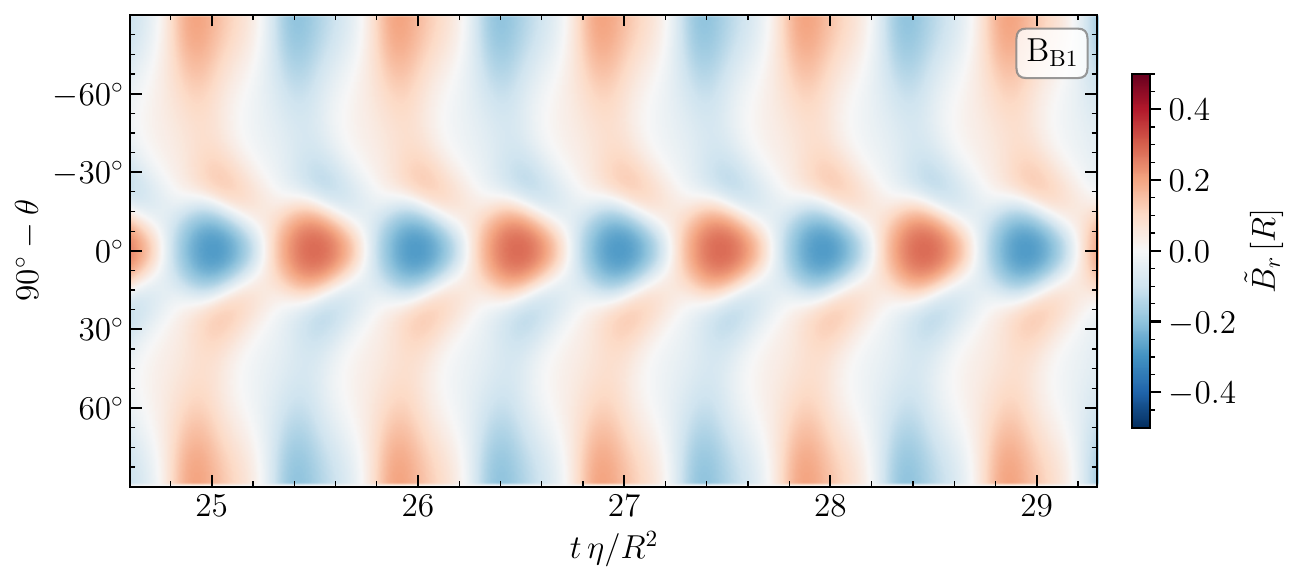}\\
    \includegraphics[width=0.33\linewidth]{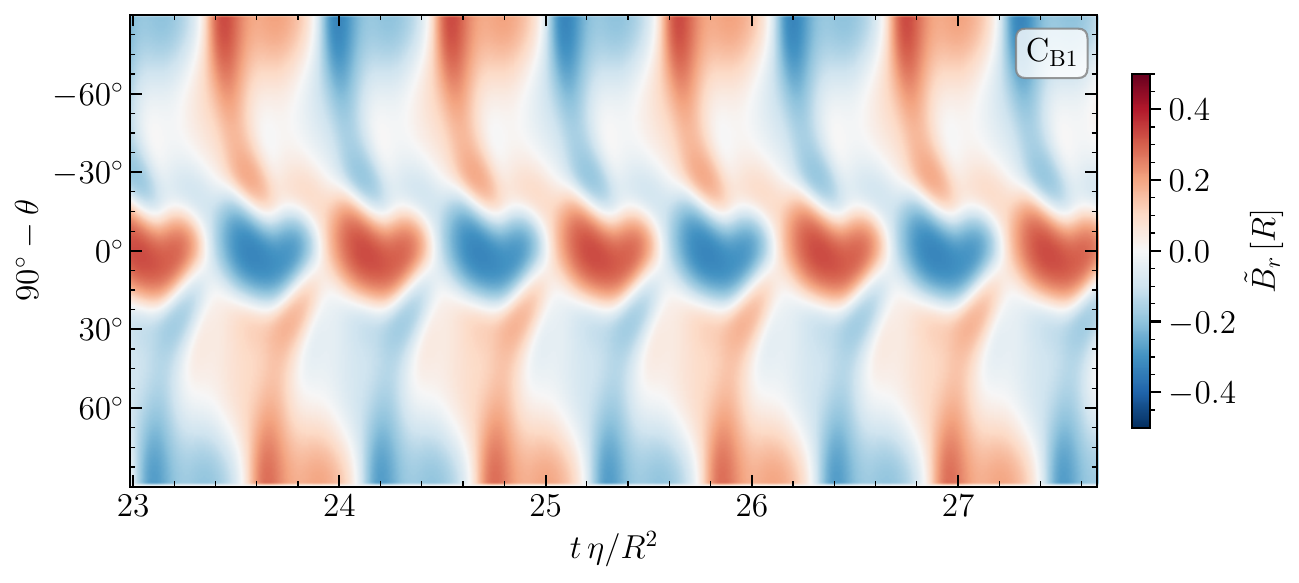}
    \includegraphics[width=0.33\linewidth]{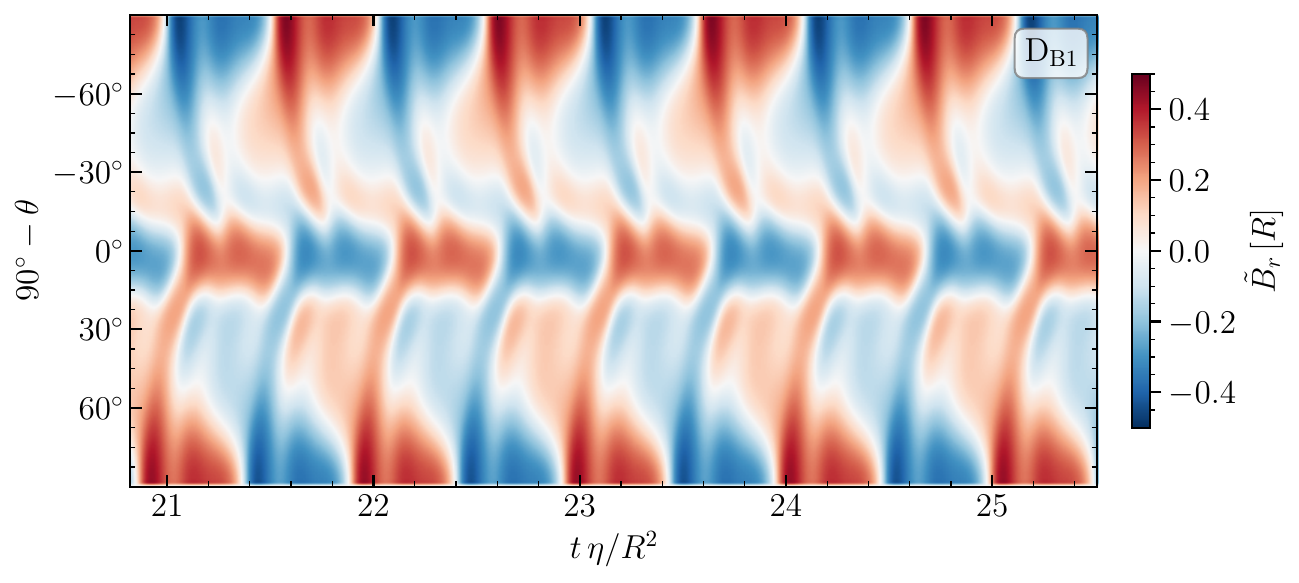}
    \includegraphics[width=0.33\linewidth]{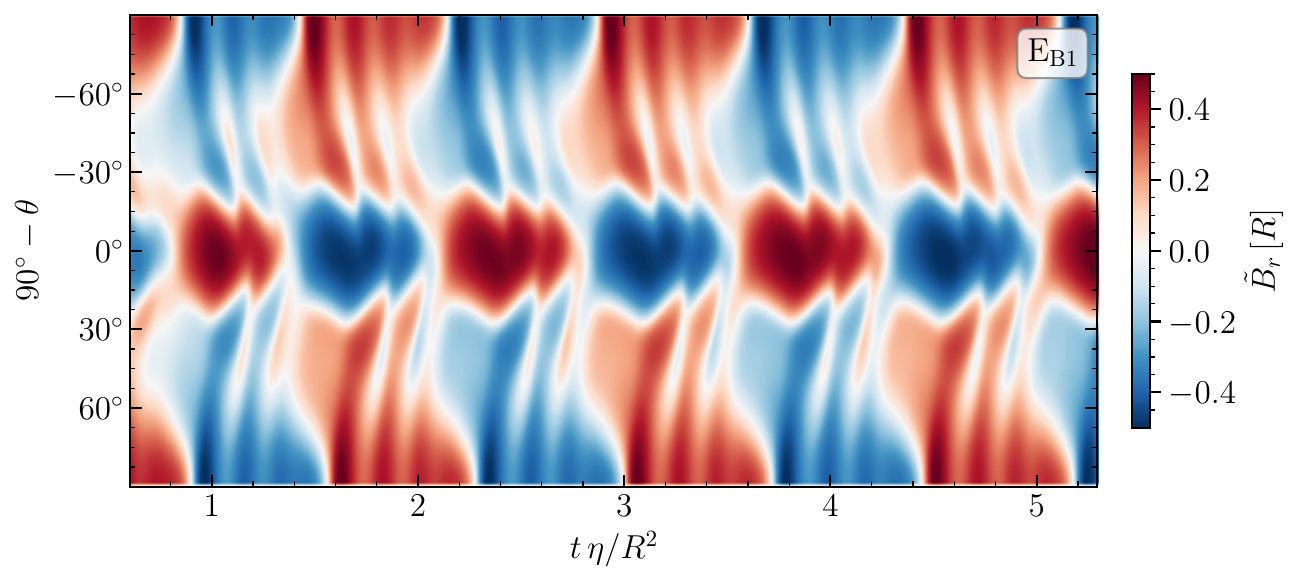}\\
    \includegraphics[width=0.33\linewidth]{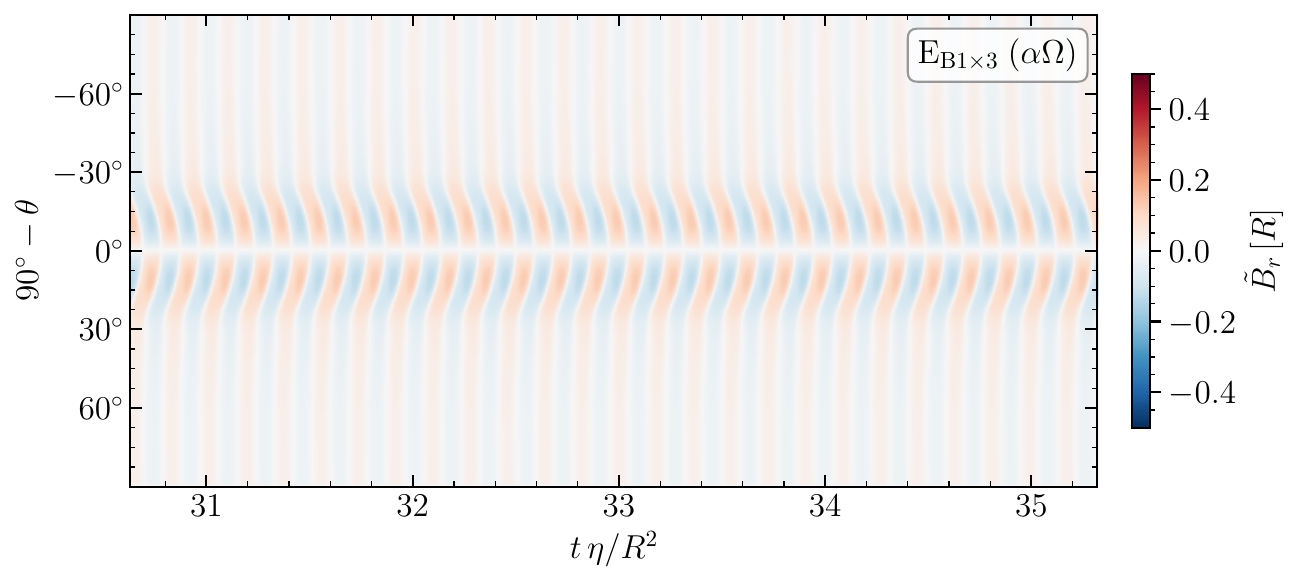}
    \includegraphics[width=0.33\linewidth]{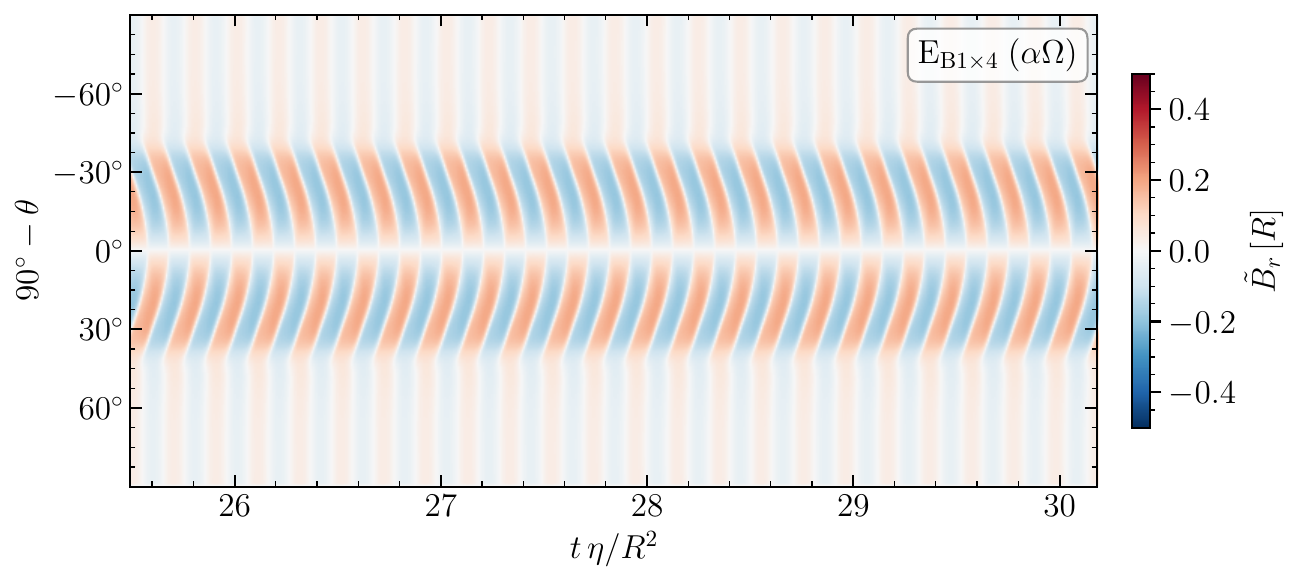}
    \includegraphics[width=0.33\linewidth]{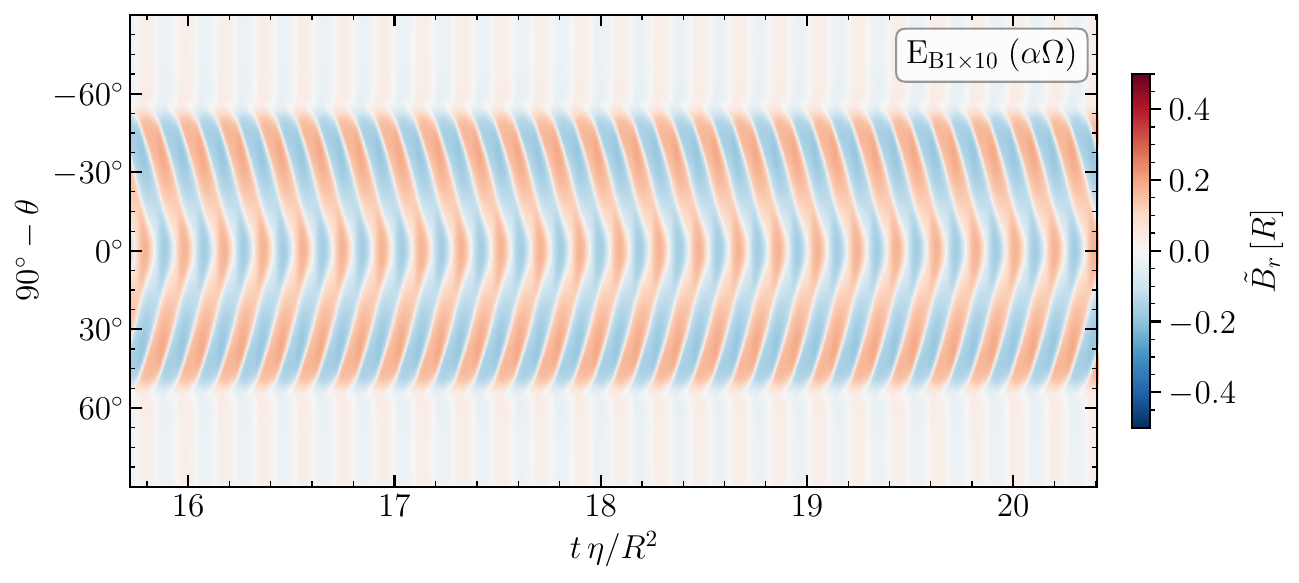}
    \caption{Butterfly (time-latitude) diagrams of the mean magnetic field component
      $\tilde{B}_r$ normalised by the equipartition field $B_{\rm eq}$
      for selected runs. The top row shows runs without differential
      rotation, $\rm{C_0}$ and $\rm{E_0}$, together with the rotating run,
      $B_{B1}$. The middle row shows the runs $\rm{C_{B1}}$, $\rm{D_{B1}}$, and
      $\rm{E_{B1}}$; the last four have the same rotation profile and
      increasing the values of $C_\alpha$. The bottom row shows runs
      with the enforced $\alpha\Omega$ approximation and fixed
      $C_\alpha$ and increasing $C_\Omega$.
      }
    \label{fig:butterfly_mf}
\end{figure*}

The magnetic field of our mean-field models was run to saturation in
representative cases of each type of dynamo explored: the full
$\alpha^2\Omega$ dynamo, runs with $\alpha\Omega$ approximation, and
in $\alpha^2$ cases.
Figure~\ref{fig:butterfly_mf} shows representative butterfly diagrams 
of the radial magnetic field for selected runs.
Runs $\rm{A_0}$, $\rm{C_0}$, and $\rm{E_0}$ without differential rotation have $C_\lambda>0$ (see Appendix~\ref{sec:complete-set}),
indicating that these are successful $\alpha^2$ dynamos.
These runs exhibit a strong radial magnetic field, $\tilde{B}_r$, near 
the surface, with the strength increasing with greater values of $C_\alpha$.
The field {of runs $\rm{A_0}$ and $\rm{C_0}$}
shows a {dominant} dipolar polarity, with opposite polarities between northern
and southern hemispheres{, while run $\rm{E_0}$ shows a quadrupolar topology.}
The magnetic field of runs $\rm{B_{B1}}$, $\rm{C_{B1}}$, $\rm{D_{B1}}$, and $\rm{E_{B1}}$ are cyclic
as shown in the top right and middle panels of Fig.~\ref{fig:butterfly_mf}.
Their radial and azimuthal components are symmetric about the equator, and
they migrate equatorward at low latitudes and poleward at higher latitudes
with increasingly tilted structures as $C_\alpha$ increases.
The latitudinal component, $\tilde{B}_{\theta}$, is antisymmetric about 
the equator. In run $\rm{B_{B1}}$, $\tilde{B}_{\theta}$ migrates equatorward,
while for higher values of $C_\alpha$ the migration becomes poleward.

The $\alpha\Omega$ models show cyclic 
reversals of the magnetic field near the surface.
In most cases, the radial and azimuthal components are 
antisymmetric with respect to the equator, while $\tilde{B}_{\theta}$ 
is symmetric,
corresponding to a predominantly dipolar parity.
The magnetic field is generally concentrated near the 
equator and shows equatorward migration, while stronger
differential rotation broadens the latitudinal extent of the field,
as shown in the bottom panel of Fig.~\ref{fig:butterfly_mf}.
Runs $\rm{C_{B1\times 10}(\alpha\Omega)}$ and  $\rm{E_{B1\times 10}(\alpha\Omega)}$ 
exhibit a transition to a quadrupolar parity. In these cases, the
radial and azimuthal components become symmetric about the equator,
while $\tilde{B}_{\theta}$ becomes antisymmetric.

\subsection{Comparing 3D and mean-field approaches}
Having established the dynamo behaviour of the mean-field models, 
we now compare them with the corresponding 3D simulations.
We compare in terms of growth rates and large-scale magnetic morphology.
Table~\ref{tab:mf_3d_comparison} summarises the main dynamo characteristics
obtained in both approaches, including cyclicity and large-scale field 
structure.
By varying $C_\alpha$ and $C_\Omega$ in the mean-field models, we obtain growth 
rates comparable to those of the 3D simulations. We select mean-field runs 
with $C_\lambda$ values within $20\%$ of the corresponding 3D values.
In general, the growth rates of the slowly rotating 3D simulations are matched by 
the mean-field runs with moderate values of $C_\alpha$ without
requiring strong shear, whereas higher rotating cases generally 
require enhanced shear or higher $C_\alpha$.
The dimensionless growth rate
of the run B1 is $C_\lambda=0.96$, which is similar to the value of the 
mean-field runs $\rm{B_{B1}}$ and $\rm{E_{B1\times3}(\alpha\Omega}$).
Both the 3D run B1 and $\rm{E_{B1\times3}(\alpha\Omega)}$ exhibit cyclic
behaviour near the surface, with radial and azimuthal components of opposite
polarities with respect to the equator (dipolar field). 
However, the latitudinal extent of 
$\rm{E_{B1\times3}(\alpha\Omega)}$
is considerably reduced in comparison to the 3D case 
(see Figs.~\ref{fig:butterfly3D} and~\ref{fig:butterfly_mf}).
By contrast, the radial magnetic field of the mean-field run 
$\rm{B_{B1}}$ shows mean quadrupolar magnetic field.
Run B2 with $C_\lambda = 0.70$, has no mean-field run counterpart with
similar growth rate. The growth rate ($\rm{C_\lambda=6.30}$) of run B5
is similar to the mean-field cases $\rm{A_0}$, $\rm{D_{B1\times5}(\alpha\Omega)}$, and 
$\rm{E_{B1\times4} (\alpha\Omega)}$. As shown in the middle panel of
Fig.~\ref{fig:butterfly3D}, the azimuthal component of B5 is, in general, symmetric
about the equator, while for $\rm{D_{B1\times5}(\alpha\Omega)}$, and 
$\rm{E_{B1\times4} (\alpha\Omega)}$ is the opposite. 
Runs $\rm{B_{B1}}$, $\rm{C_{B1}}$, $\rm{D_{B1}}$, and $\rm{E_{B1}}$ produce large-scale
magnetic field extending from the equator to the poles, similarly to
what is observed in run B5.

The fastest rotating cases, B10 and B20, have $C_\lambda$ values of 
12.22 and 11.97, respectively. Two mean-field runs have comparable
growth rates ($\rm{A_{B1\times10}(\alpha\Omega)}$ and $\rm{C_{B1}}$), but
the rapidly rotating 3D runs show a transition to a non-axisymmetric
large-scale magnetic field.
Since our mean-field models are axisymmetric, we cannot model the
predominantly non-axisymmetric fields of the rapidly rotating 3D runs.

The values of $C_\alpha$ in the mean-field runs are very similar to those 
of the 3D simulations (see Tables~\ref{tab:dynamo_numbers} and \ref{tab:mf_full}),
as $C_\alpha$ was varied during the parameter study to obtain growth rates comparable
to those of the 3D runs. However, comparable growth rates
are obtained only for some combinations of $C_\alpha$ and $C_\Omega$, and matching the
growth rate does not uniquely determine the resulting large-scale magnetic morphology.

\begin{table*}
\centering
\caption{Comparison of the dynamo behaviour between the 3D and the mean-field simulations.}
\label{tab:mf_3d_comparison}
\begin{tabular*}{\textwidth}{@{\extracolsep{\fill}} C{5cm} C{5cm} C{8cm}}
\hline\hline
Approach / run family & Dynamo type & Cyclic behaviour / morphology \\
\hline

3D: B1--B2
& Axisymmetric large-scale dynamo
& Cyclic; predominantly dipolar; equatorward  \\

3D: B5
& Axisymmetric large-scale dynamo; transitional regime
& Cyclic; quadrupolar contribution near the surface; equatorward \\

3D: B10--B20
& Non-axisymmetric; Rapid-rotation dynamo; rotationally constrained
& No cycles; possible azimuthal dynamo waves \\

\hline

MF: full models, $\Omega=0$
& $\alpha^2$
& Non cyclic; growing fields without shear; dipolar radial and latitudinal components {for low enough $C_{\alphaß}$}\\

MF: full models, $\Omega_{\rm B1}$
& $\alpha^2\Omega$
& Cyclic; symmetric about the equator; equatorward (poleward) at mid (high) latitudes \\

MF: enforced $\alpha\Omega$, weak shear
& Subcritical $\alpha\Omega$
& No dynamo \\

MF: enforced $\alpha\Omega$, strong shear
& Supercritical $\alpha\Omega$
& Cyclic; equatorward migration; transition to quadrupolar parity \\

\hline\hline
\end{tabular*}
\end{table*}

\section{Conclusions}
We have studied the origin of magnetic fields in
{the envelopes of} red giants using the results
of 3D MHD simulations to constrain mean-field parameters for 2D
mean-field models. Our results indicate that even $\alpha^2$
mean-field models without shear lead to dynamo action.
This is interesting in view of the puzzling observations of
\cite{tsvetkova2013magnetic} that the giant $\beta$ Ceti has a strong
dipolar surface magnetic field, but no surface differential rotation
and a long rotation period of about 215 days, corresponding to a 
Rossby number of about 1.3.
This star may therefore not be the descendant of an Ap star, as suggested
by the authors, but the magnetic field could be explained by an
$\alpha^2$ dynamo.
{Similarly, spectropolarimetric observations of the slow-rotating 
star Pollux by \cite{auriere2021pollux}, reveal a dominant poloidal 
magnetic field component with dipolar topology, pointing to the
possible role of turbulence for dynamo generation \citep{lehtinen2020common}.
This is in agreement with our rigidly rotating mean-field runs ${\rm A_0}$ and 
${\rm C_0}$, where an $\alpha^2$ dynamo is at work.
In terms of surface magnetic field strengths, our 3D runs 
($\sim 83$ -- $320$~G)
are within one order of magnitude higher than reported strengths
for $\beta$ Ceti (20~G) \citep{tsvetkova2013magnetic} or V390 Aurigae (14~G) 
\citep{konstantinova2012magnetic}. Since these measurements correspond to the
large-scale magnetic field, they may underestimate the total field strength,
so the difference with our 3D simulations could be smaller.
}

Differential rotation
changes the growth rate of the magnetic field such that it decreases
it in runs with low $C_\alpha$ and increases it for sufficiently high
$C_\alpha$, whereas $\alpha\Omega$ dynamos require sufficiently strong
shear to become supercritical.
{Furthermore, our rapidly rotating 3D runs show less clear 
axisymmetric field structures, suggesting a transition towards 
non-axisymmetric modes. This is consistent with the behaviour of
the fast rotator V390 Aur, which exhibits a prominent high-latitude 
toroidal belt and a complex surface topology \citep{konstantinova2012magnetic}.
}
Regarding the survival of a fossil magnetic field, we find that,
in the parameter range we explored, differential rotation without
$\alpha$ effect does not prolong the magnetic decay time.
This supports the interpretation that the sustained magnetic fields 
{in the envelopes of} red giants require dynamo action.

In general, the mean-field models can reproduce the 3D growth rates, but the 
correspondence is not unique, since different combinations of $C_\alpha$
and rotation profile can give similar $C_\lambda$. Therefore, the growth
rate alone is not enough to compare consistently among the two approaches.
Furthermore, the mean-field models reveal that, despite having similar
growth rates, the dynamo mode can be different: both full and
$\alpha\Omega$ models can yield similar values of $C_\lambda$, and exhibit 
different large-scale magnetic field configurations.
The mean-field models reproduce several qualitative features of the slowly
and moderately rotating 3D runs, including cyclic behaviour and equatorward
migration of the large-scale magnetic field, but differences 
remain in the parity and the latitudinal extent of the magnetic field.
The 3D simulations show that faster rotation rates enhance the dynamo 
efficiency, although the rotational dependence of the growth of the
magnetic field is not completely monotonic. For example, the fields in
run B2 grow slower than in run B1. Furthermore, the more rapidly rotating
cases show a transition away from predominantly axisymmetric large-scale
fields toward a dominating non-axisymmetric $m=1$ mode. Therefore,
increasing rotation does not only strengthen the dynamo, but it also changes
its mode of operation.
{Our results can be compared with the work of \cite{noraz2022impact}
with mean field dynamos for solar like stars. They concluded that traditional
turbulent dynamos can sustain cycles for antisolar differential 
rotation while Babcock-Leighton models in general do not. 
We consider only turbulent dynamos and also find cyclic solutions
in our mean-field models, in agreement with our 3D simulations.
}

The 3D and mean-field runs only occupy negative dynamo numbers $D$ in the R\"adler diagram.
This is because the anisotropy of the turbulence is such that it leads
to a negative radial angular velocity gradient.
As alluded to in the introduction, this is consistent with 3D anelastic
hydrodynamical simulations of rotating red giant convection by
\cite{Brun2009}.
While this result is naturally explained in terms of hydrodynamic
mean-field theory using the $\Lambda$ effect \citep{Ruediger80},
in the context of giants, a
negative radial angular velocity gradient is traditionally explained as
the result of a rapidly rotating core that, unlike the envelope of the
star, did not yet have enough time to spin down.
Given that simulations now point to a naturally occurring negative angular
velocity gradient as a consequence of anisotropic convection, it casts
doubt on earlier ideas about a rapidly spinning core at its formation
as the only reason for it. In our simulation B1, the core is rotating at 
least twice as fast as the envelope, while 
\cite{li2024asteroseismic} reported 
core-to-envelope rotation ratios that vary
by a factor of two for a subset of their sample. A deeper understanding of the 
angular momentum transport mechanism in red giant stars is therefore still needed
\citep{aerts2019angular, cantiello_angular_2014}.

\section*{Data availability}
The simulations presented in this work were performed using the \textsc{Pencil Code}
\citep{brandenburg2021pencil}, which is available at
\url{https://github.com/pencil-code/}. The corresponding local input files 
and source local configuration files are publicly available at 
\url{https://doi.org/10.5281/zenodo.21361381}.

\begin{acknowledgements}
COR acknowledges financial support from ANID (ANID DOCTORADO DAAD-BECAS
CHILE/62220030) as well as financial support from DAAD (DAAD/BECAS
Chile funding program ID 57636841).
The simulations were performed with
resources provided by the NHR@ZIB cluster under project grant hhp00064, and the Kultrun Astronomy Hybrid Cluster via the
projects Conicyt Quimal \#170001, Anillo ACT172033, Fondecyt regular 1180291,
Fondecyt Iniciacion 11170268, Basal AFB-170002, and Núcleo Milenio 
Titans NCN19-058. COR thanks Matthias Rheinhardt for helpful
discussions and assistance with code development and debugging. 
Part of this work was carried out during a research visit at Nordita
supported by the Nordita Visiting PhD Fellow Programme.
This research was supported in part by the
Swedish Research Council (Vetenskapsr{\aa}det) under Grant No.\ 2025-05957.
PJK acknowledges financial support from the European Research Council through the ERC Advanced Grant NEOCON under grant No. 101199531.
\end{acknowledgements}

\bibliographystyle{aa}
\bibliography{paper}

\begin{appendix}
\onecolumn

\section{Table with the complete set of mean-field runs}
\label{sec:complete-set}
In Sect.~\ref{sec:Mean-field-model}, we presented our mean-field models.
Here we present a list with the complete set of mean-field runs; see Table~\ref{tab:mf_full}.

\begin{table}[h!]
\caption{Complete set of mean-field simulations. Runs are labelled by the
imposed rotation profile $\Omega$ and the value of $C_\alpha$. Models with 
enforced $\alpha\Omega$ approximation are indicated explicitly. The table 
lists the dimensionless growth rate $C_\lambda$, the
shear $\Delta\Omega$, and the corresponding $C_\Omega$.
}
\label{tab:mf_full}
\centering
\begin{tabular}{c c c c c c}
\hline\hline
Run & $\Omega$ & $C_\lambda$ & $\Delta\Omega$ & $C_\Omega$ & $B_{\rm rms}/B_{\rm eq}$\\
\hline
\multicolumn{6}{c}{\it $C_\alpha = 1.90$ {\rm (Set A)}} \\
\hline
${\rm{A_{0}}}$ & --  & $5.13$ &  -- & -- & 0.14\\
${\rm{A_{B1}}}$ & $\Omega_{\rm B1}$ & $-0.56$ & $-1.35$ & $-7.29$ & --\\
${\rm{A_{B1}({\alpha\Omega})}}$ & $\Omega_{\rm B1}$ & $-4.79$ &  $-1.35$ & $-7.29$ &--\\
${\rm{A_{B1\times2}({\alpha\Omega})}}$ & $2\Omega_{\rm B1}$ & $-5.28$ & $-2.70$ & $-14.6$ &--\\
${\rm{A_{B1\times3}({\alpha\Omega})}}$ & $3\Omega_{\rm B1}$ & $-3.20$ & $-4.05$ & $-21.9$ &--\\
${\rm{A_{B1\times4}({\alpha\Omega})}}$ & $4\Omega_{\rm B1}$ & $-0.70$ & $-5.40$ & $-29.2$ &--\\
${\rm{A_{B1\times5}({\alpha\Omega})}}$ & $5\Omega_{\rm B1}$ & $1.92 $ & $-6.75$ & $-36.5$ &0.71\\
${\rm{A_{B1\times10}({\alpha\Omega})}}$ & $10\Omega_{\rm B1}$ & $12.8$& $-13.50$ & $-72.9$&1.90\\
\hline
\multicolumn{6}{c}{\it $C_\alpha = 2.12$ {\rm (Set B)}} \\
\hline
${\rm{B_{B1}}}$ & $\Omega_{\rm B1}$  & $1.07$ &  $-1.35$ & $-7.29$ &0.41\\
${\rm{B_{B1}({\alpha\Omega})}}$ & $\Omega_{\rm B1}$  & $-4.92$ & $-1.35$ & $-7.29$ & --\\
${\rm{B_{B1\times2}({\alpha\Omega}}}$) & $2\Omega_{\rm B1}$ & $-4.87$ &  $-2.70$ & $-14.6$& --\\
${\rm{B_{B1\times3}({\alpha\Omega}}}$) & $3\Omega_{\rm B1}$ & $-2.40$ &  $-4.05$ & $-21.9$& --\\
${\rm{B_{B1\times4}({\alpha\Omega}}}$) & $4\Omega_{\rm B1}$ & $0.47 $ &  $-5.40$ & $-29.2$&0.22\\
${\rm{B_{B1\times5}({\alpha\Omega}}}$) & $5\Omega_{\rm B1}$ & $3.34 $ &  $-6.75$ & $-36.5$&0.94\\
\hline
\multicolumn{6}{c}{\it $C_\alpha = 2.33$ {\rm (Set C)}} \\
\hline
${\rm{C_{0}}}$ & -- & $15.7$ &  -- & -- & 0.27\\
${\rm{C_{B1}}}$ & $\Omega_{\rm B1}$  &$9.96$ &  $-1.35$ & $-7.29$ & 0.57 \\
${\rm{C_{B1}({\alpha\Omega})}}$ & $\Omega_{\rm B1}$  & $-5.12$ & $-1.35$ & $-7.29$ & --\\
${\rm{C_{B1\times2}({\alpha\Omega})}}$ & $2\Omega_{\rm B1}$ & $-4.42$ & $-2.70$ & $-14.6$ &--\\
${\rm{C_{B1\times3}({\alpha\Omega})}}$ & $3\Omega_{\rm B1}$ & $-1.56$ & $-4.05$ & $-21.9$ &--\\
${\rm{C_{B1\times4}({\alpha\Omega})}}$ & $4\Omega_{\rm B1}$ & $1.63$  & $-5.40$ & $-29.2$ &0.64\\
${\rm{C_{B1\times5}({\alpha\Omega})}}$ & $5\Omega_{\rm B1}$ & $4.73$  & $-6.75$ & $-36.5$ &1.10\\
${\rm{C_{B1\times10}({\alpha\Omega})}}$& $10\Omega_{\rm B1}$& $17.0$  &  $-13.50$ & $-72.9$ & 2.38\\
\hline
\multicolumn{6}{c}{\it $C_\alpha =  2.76$ {\rm (Set D)}} \\
\hline
${\rm{D_{B1}}}$ & $\Omega_{\rm B1}$  & $24.8$ &  $-1.35$ & $-7.29$ & 0.64\\
${\rm{D_{B1}}} ({\alpha\Omega})$ & $\Omega_{\rm B1}$  & $-5.15$ &  $-1.35$ & $-7.29$ & --\\
${\rm{D_{B1\times2} ({\alpha\Omega})}}$ & $2\Omega_{\rm B1}$ & $-3.47$ & $-2.70$ & $-14.6$ & --\\
${\rm{D_{B1\times3} ({\alpha\Omega})}}$ & $3\Omega_{\rm B1}$ & $0.18 $ & $-4.05$ & $-21.9$ & --\\
${\rm{D_{B1\times4} ({\alpha\Omega})}}$ & $4\Omega_{\rm B1}$ & $3.89 $ & $-5.40$ & $-29.2$ & 1.00\\
${\rm{D_{B1\times5} ({\alpha\Omega})}}$ & $5\Omega_{\rm B1}$ & $7.32 $ & $-6.75$ & $-36.5$ & 1.37\\
\hline
\multicolumn{6}{c}{\it $C_\alpha = 2.97$ {\rm (Set E)}} \\
\hline
${\rm{E_{0}}}$                         & --             & $21.9$  & $-$& $-$ &0.43 \\
${\rm{E_{0} (\alpha\Omega)}}$          & --             & $-2.02$ & $-$ & $-$ & --\\
${\rm{E_{B1}}}$                        & $\Omega_{\rm B1}$& $27.2$  & $-1.35$ & $-7.29$ & 1.03\\
${\rm{E_{B1\times2}}}$                 & $2\Omega_{\rm B1}$    & $30.2$  & $-2.70$ & $-14.6$ & 1.17\\
${\rm{E_{B1} ({\alpha\Omega})}}$       & $\Omega_{\rm B1}$& $-5.40$ & $-1.35$ & $-7.3$ &-- \\
${\rm{E_{B1\times2}({\alpha\Omega})}}$ & $2\Omega_{\rm B1}$    & $-2.94$ & $-2.70$ & $-14.6$ &--\\
${\rm{E_{B1\times3}({\alpha\Omega})}}$ & $3\Omega_{\rm B1}$    & $ 1.05$ & $-4.05$ & $-21.9$ & 0.42\\
${\rm{E_{B1\times4}({\alpha\Omega})}}$ & $4\Omega_{\rm B1}$    & $ 5.03$ & $-5.40$ & $-29.2$ & 1.12\\
${\rm{E_{B1\times5}({\alpha\Omega})}}$ & $5\Omega_{\rm B1}$    & $8.69$  & $-6.75$ & $-36.5$ & 1.48\\
${\rm{E_{B1\times10}({\alpha\Omega})}}$& $10\Omega_{\rm B1}$  & $22.9$   & $-13.50$ & $-72.9$& 2.91\\
${\rm{E_{B10}({\alpha\Omega})}}$       & $\Omega_{\rm B10}$            & $-1.01$  & $-0.84$ & $-4.51$ & --\\

\hline
\multicolumn{6}{c}{\it $C_\alpha = 0$ {\rm (Set F)}} \\
\hline
${\rm{F_{0}}}$                         & --            & $-1.98$ & -- & -- & --\\
${\rm{F_{B1} (\alpha\Omega)}}$         & $\Omega_{\rm B1}$    & $-2.28$ &  $-1.35$ & $-7.29$ &--\\
${\rm{F_{B1\times2}({\alpha\Omega})}}$ & $2\Omega_{\rm B1}$   & $-2.09$ & $-2.70$ & $-14.6$  &--\\
${\rm{F_{B1\times3}({\alpha\Omega})}}$ & $3\Omega_{\rm B1}$   & $-2.05$ & $-4.05$ & $-21.9$  &--\\
${\rm{F_{B1\times4}({\alpha\Omega})}}$ & $4\Omega_{\rm B1}$   & $-2.04$ & $-5.40$ & $-29.2$  &--\\
${\rm{F_{B1\times5}({\alpha\Omega})}}$ & $5\Omega_{\rm B1}$   & $-2.03$ & $-6.75$ & $-36.5$  &--\\
${\rm{F_{B1\times10}({\alpha\Omega})}}$& $10\Omega_{\rm B1}$  & $-2.02$ & $-13.50$ & $-72.9$ &--\\
\hline \hline
\end{tabular}
\end{table}

\twocolumn
\section{Gegenbauer polynomials}
\label{sec:gegenbauer}
\begin{figure}[h!]
\centering
\includegraphics[width=\columnwidth]{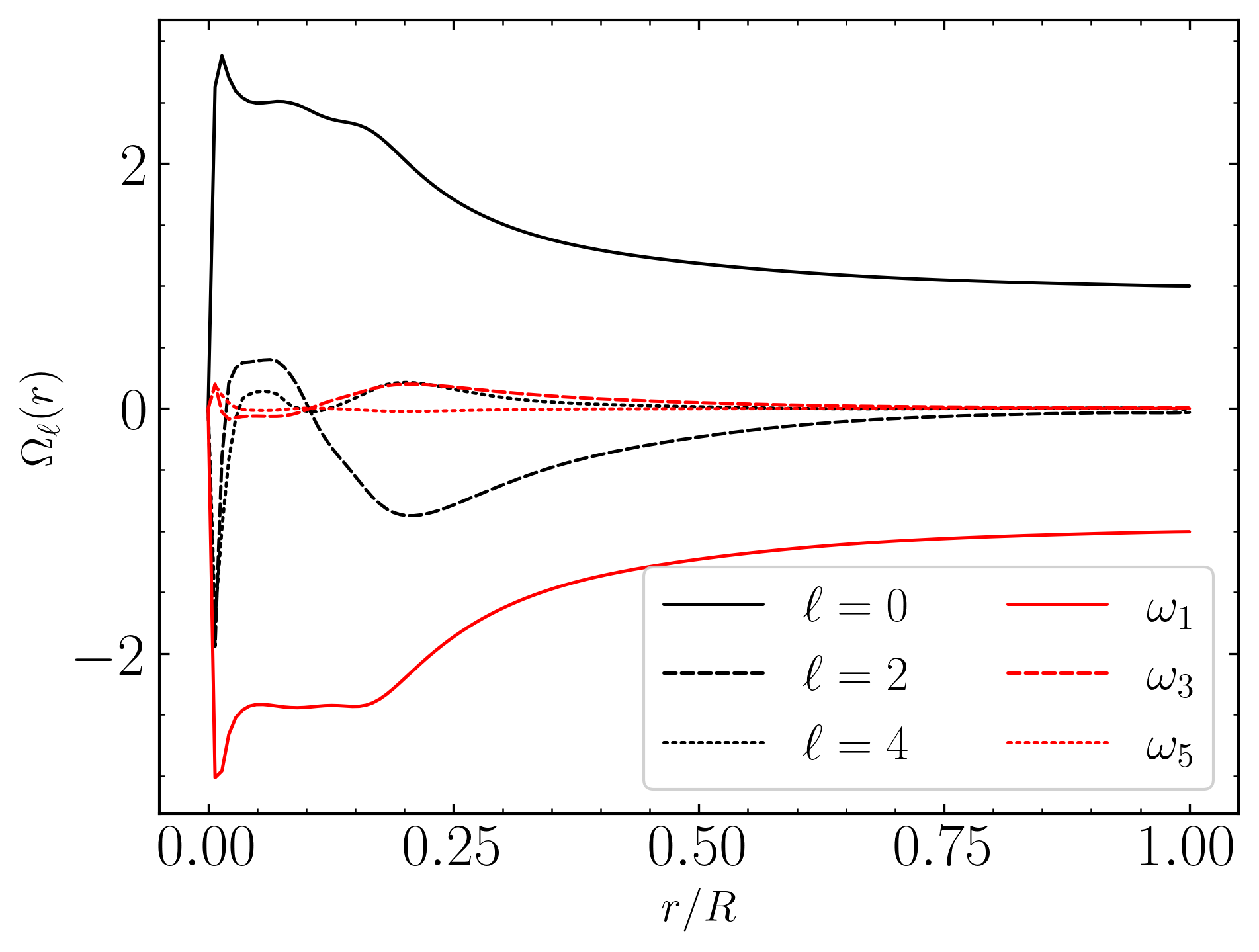}
\caption[]{Radial dependence of the first three even Legendre coefficients $\Omega_\ell(r)$ $(\ell=0,2,4\text{, black})$ together with the corresponding Gegenbauer coefficients $\omega_1$, $\omega_3$, and $\omega_5$ (red) for run B1.} 
\label{pgegenbauer}
\end{figure}

In Sect.~\ref{sec:Mean-field-model}, we expanded the angular velocity
in terms of Legendre polynomials.
A different expansion is in terms of Gegenbauer polynomials.
Figure~\ref{pgegenbauer} compares the radial dependence of the first three 
even Legendre modes retained in the present work with the corresponding 
Gegenbauer functions. The Gegenbauer representation concentrates most
of the structure in the lowest-order modes and therefore converges 
more rapidly.
However, for a fixed truncation level, the two expressions are equivalent.
Nevertheless, the Legendre modes used in this 
work capture the large-scale structure of the rotation profile 
of the 3D run B1.

\section{Growth rate vs.\ dynamo number}
\label{Growth-rate-vs-dynamo-number}
\begin{figure}[h!]\begin{center}
\includegraphics[width=\columnwidth]{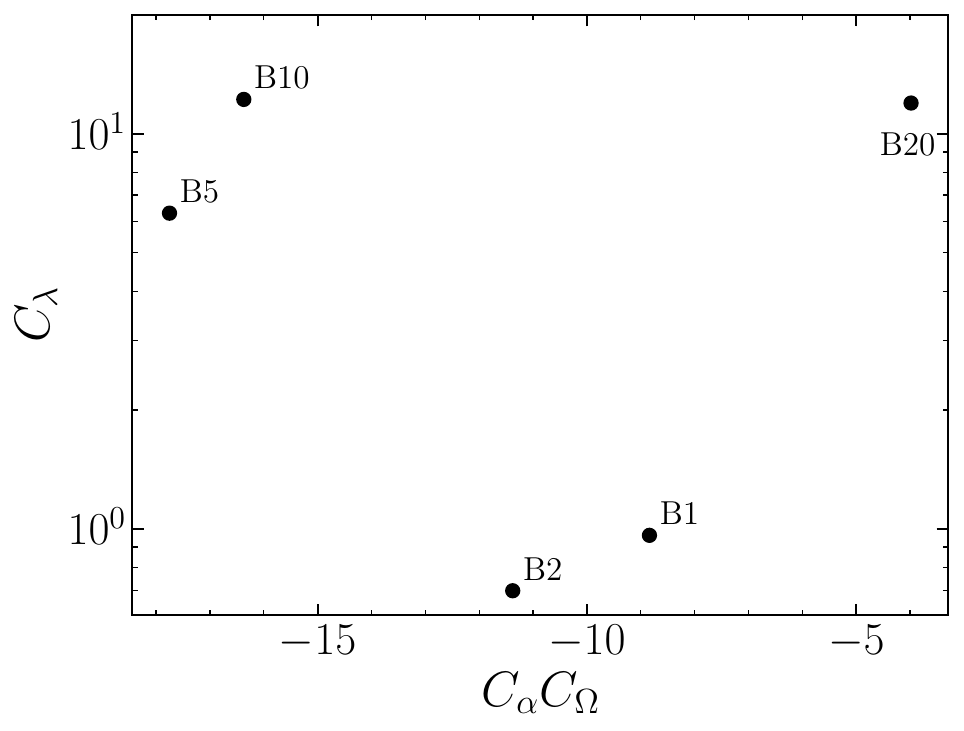}
\end{center}
\caption[]{$C_{\lambda}$ as function of the dynamo number $C_{\alpha}C_{\Omega}$ for the 3D
simulations. 
}
\label{fig:ClambdavsCoCa}
\end{figure}

In Sect.~\ref{sec:Mean-field-parameter-study}, we presented 
a mean-field parameter study and discussed the dependence
of $C_{\lambda}$ on the dynamo number $C_{\alpha}C_{\Omega}$ for the 3D simulations.
Figure~\ref{fig:ClambdavsCoCa} shows the normalised growth rate $C_{\lambda}$
as a function of the product $C_\alpha C_\Omega$, for the 3D simulations.
The dynamo efficiency does not follow a simple dependence on the dynamo number
across all the rotation rates considered here, particularly for the rapidly rotating
cases. 
\end{appendix}
\end{document}